\documentclass[aps,prb,amsmath ,amssymb, twocolumn, superscriptaddress, longbibliography]{revtex4-2}

\usepackage{mathtools,amssymb,graphicx,units,bm}
\usepackage{amsmath}
\usepackage[plainpages=false,pdfpagelabels,colorlinks=true,linkcolor=PineGreen,urlcolor=PineGreen,citecolor=PineGreen,pdftitle={Title},pdfauthor={},pdfdisplaydoctitle=true,pdfduplex=DuplexFlipLongEdge]{hyperref}
\usepackage{soul} 
\usepackage[dvipsnames,usenames]{color}
\usepackage[normalem]{ulem}
\graphicspath{{figures/}}
\usepackage{color}
\usepackage{bbold} 
\usepackage{float}
\usepackage{orcidlink}
\emergencystretch=\maxdimen
\usepackage{dcolumn}
\usepackage{bm}
\usepackage{physics}%
\usepackage[dvipsnames]{xcolor}
\usepackage{bbold} 
\usepackage{float}
\emergencystretch=\maxdimen
\def\s{\sigma}
\def\ave#1{\langle #1\rangle}
\def\ket#1{\left \vert #1\right \rangle}

\newcommand{\iv}{\mathbf{i}}
\newcommand{\jv}{\mathbf{j}}
\newcommand{\qv}{\mathbf{q}}
\newcommand{\rv}{\mathbf{r}}
\newcommand{\kv}{\mathbf{k}} 

\newcommand{\av}{\mathbf{a}}

\usepackage[normalem]{ulem}

\begin{document}

\normalem

\title{Unconstrained $\mathbb{Z}_2$ gauge theory on a triangular lattice:  deconfined fermions in a vison crystal and superfluidity}

\author{Welberth Kennedy\, \orcidlink{0000-0001-7297-4080}}
\email{welberth@pos.if.ufrj.br}
\affiliation{Instituto de F\'isica, Universidade Federal do Rio de Janeiro Cx.P. 68.528, 21941-972 Rio de Janeiro RJ, Brazil}
\affiliation{\mbox{Institut f\"ur Theoretische Physik und Astrophysik,
		Universit\"at W\"urzburg, 97074 W\"urzburg, Germany}}
\author{Jo\~ao C. In\'acio\, \orcidlink{0009-0006-5457-3711}}
\email{joao.carvalho-inacio@uni-wuerzburg.de}
\affiliation{\mbox{Institut f\"ur Theoretische Physik und Astrophysik,
		Universit\"at W\"urzburg, 97074 W\"urzburg, Germany}}
\author{Fakher F. Assaad\, \orcidlink{0000-0002-3302-9243}}
\affiliation{\mbox{Institut f\"ur Theoretische Physik und Astrophysik,
    Universit\"at W\"urzburg, 97074 W\"urzburg, Germany}}
\affiliation{W\"urzburg-Dresden Cluster of Excellence ctd.qmat, Am Hubland, 97074 W\"urzburg, Germany}

\begin{abstract}
We investigate an unconstrained $\mathbb{Z}_2$ gauge theory coupled to fermions on the triangular lattice at half filling, using auxiliary-field quantum Monte Carlo simulations. The triangular geometry gives rise to a global $\mathbb{Z}_2$ symmetry that acts as an Ising symmetry on the plaquette fluxes. At weak transverse field $h$, this symmetry is spontaneously broken, stabilizing a vison crystal. We argue that this phase is deconfined: fermions, charged under an emergent $\mathbb{Z}_2$ gauge field, propagate in an ordered $\pi$-flux background and form a Dirac semimetal. Increasing the transverse field triggers a first-order quantum phase transition to a confined $s$-wave superfluid, at which confinement and global symmetry breaking occur simultaneously. This contrasts with the square lattice, where the corresponding transition is continuous. We extend our analysis to finite temperature and determine the thermal phase boundaries of both ordered phases. The superfluid melts through a Kosterlitz--Thouless transition, whereas the vison crystal undergoes a thermal Ising transition. Together, our results establish the $T \times h$ phase diagram and demonstrate how the triangular geometry leads to qualitative changes in both the phases and the criticality of the model.

\end{abstract}


    \maketitle


\section{Introduction}
\label{sec:introduction} 

Gauge theories play an important role in our understanding of phases of matter in the solid state. They very naturally emerge in parton constructions of spin \cite{Affleck88,Affleck88a,wen_quantum_2002}, Kondo lattice \cite{Coleman84,Coleman87,Pan25} and $t$-$J$ models \cite{Lee06_rev}. In this approach the local emergent gauge symmetry stems from the redundancy of the parton representation \cite{Wen_book}. The Kitaev spin model is a canonical example since the parton construction in terms of Majorana fermions paves the way to an exact solution \cite{Kitaev06}. 
Parton constructions are not the only route to gauge theories. For instance, spin ice with its two-in/two-out constraint -- corresponding to the Gauss law -- is a classical example of a U(1) gauge theory \cite{Castelnovo08} hosting monopoles and emergent photons. As already hinted, gauge theories provide a framework to understand phases of matter. This includes spin liquids, broadly defined as a deconfined phase of an emergent gauge theory \cite{Balents10}, the heavy fermion phase in Kondo lattice models as a Higgs phase of a U(1) gauge theory \cite{Coleman84,Coleman87,Raczkowski22}, and Kondo destruction \cite{Senthil03}.

\begin{figure}[t]
    \centering
    \includegraphics[scale=0.4]{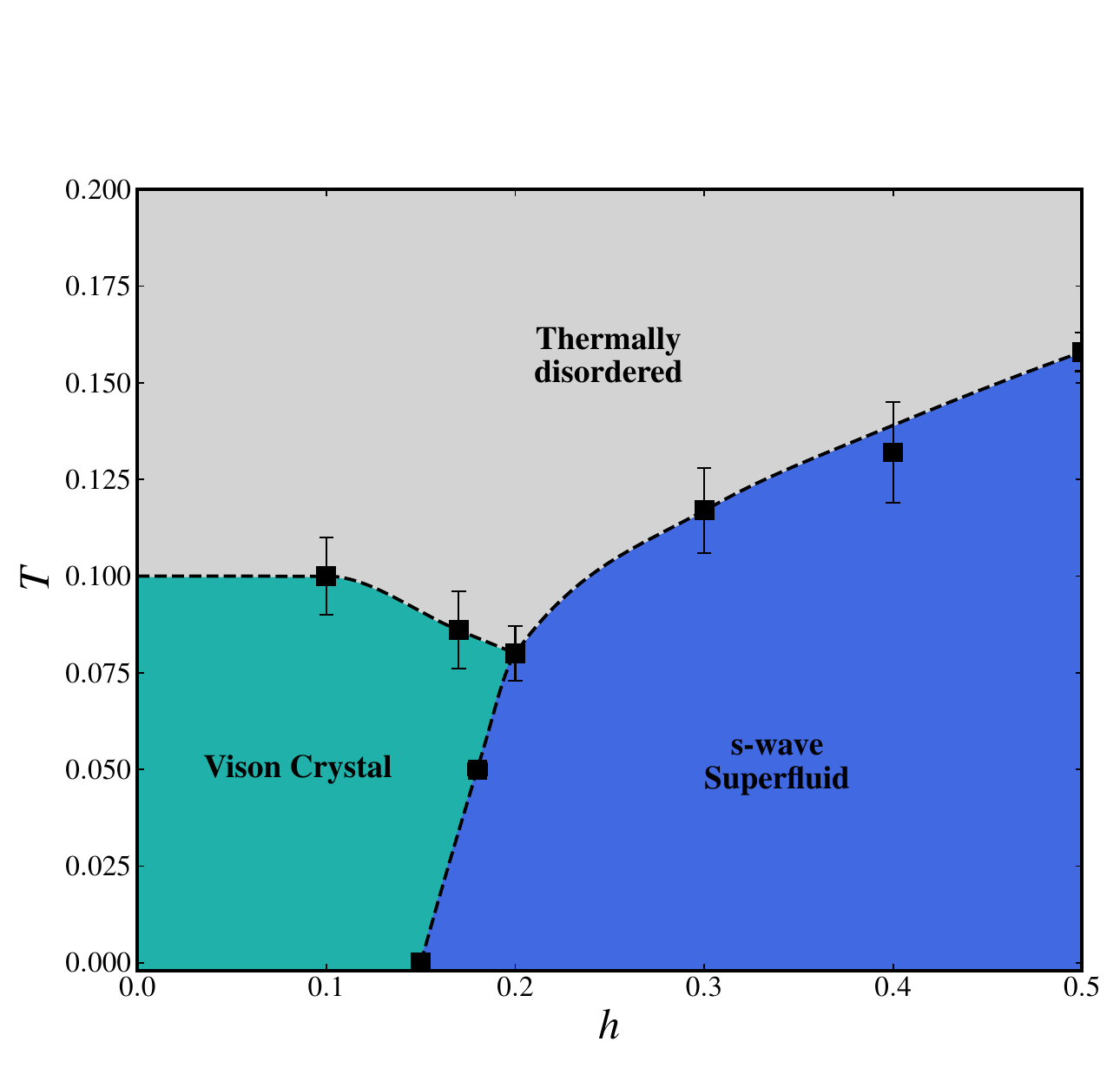}
    \caption{$T \times h$ phase diagram of the unconstrained $\mathbb{Z}_2$ gauge theory on the triangular lattice at half filling, Eq.~\eqref{eq:hamiltonian}. Black symbols indicate QMC results, while dashed lines are just guides to the eye. When not visible, error bars are smaller than the markers. The transition between the SF and the vison crystal phases is first-order as both phases break distinct symmetries: a continuous global $U(1)$ symmetry in the SF phase and a discrete global $\mathbb{Z}_2$ symmetry in the vison crystal phase. The thermal transition out of the vison crystal phase belongs to the Ising universality class, while for the SF it is a Kosterlitz--Thouless transition. The slope of the coexistence line between SF and vison crystal phases is supported by a Clausius--Clapeyron relation derived for our model in Appendix \ref{App:Clausius-Clapeyron}.}
    \label{fig:diagram}
\end{figure}

Parton constructions of the Heisenberg model provide an interesting way to study gauge theories in a solid state setting as they provide an elegant description of quantum spin liquid states. In these constructions, there is an ambiguity in the choice of the gauge group used to decouple the interaction in terms of the partons. Usual constructions use the U(1) or SU(2) groups \cite{wen_quantum_2002,song_unifying_2019,hermele_algebraic_2005,wang_deconfined_2017}. The decoupling in terms of a \(\mathbb{Z}_2\) gauge field can be viewed as a Higgs descendant of the U(1) decoupling \cite{shackleton_sign-problem-free_2025}.

In this work, we study an unconstrained \(\mathbb{Z}_2\) lattice gauge theory coupled to \(N_f = 2\) fermionic matter on the triangular lattice using numerically exact auxiliary-field quantum Monte Carlo (AF-QMC) simulations. By unconstrained, we mean that the Gauss operators are locally conserved, but their eigenvalues are not fixed, such that Gauss's law will be dynamically imposed at low energies \cite{xu_monte_2019,inacio_unconstrained_2026}. 
On the square lattice, this model has been studied extensively, and shows a plethora of phases \cite{Assaad2016,Gazit16} and exotic quantum phase transitions \cite{Gazit18}.  On bipartite lattices, the model with \(N_f\) fermion flavors has a global \(\text{O}(2N_f)\) symmetry and no sign problem for all values of \(N_f\). The model is defined by a single parameter, corresponding to the strength of the transverse field $h$. In the $h$ versus $N_f$ phase diagram, we observed orthogonal semi-metallic (OSM) \cite{Nandkishore12},  
antiferromagnetic (AFM), charge density wave (CDW)  and valence bond solid (VBS) states.  Quantum phase transitions include deconfined quantum critical points at $N_f = 3$ between VBS and  AFM  orders \cite{Senthil04_2} as well as continuous and  direct quantum phase transitions between the OSM and confined AFM phases with emergent SO(5) symmetry \cite{Gazit18}. The model of Ref.~\cite{Assaad2016}  is free of the negative sign problem even on frustrated lattices.

It is known that the lattice geometry, in particular whether or not it is bipartite, has a profound impact on the phase diagrams of gauge theories. The canonical example is the quantum dimer model, which on bipartite lattices has a mapping onto height models and emergent U(1) symmetry, but on the triangular lattice provides an instance of a $\mathbb{Z}_2$ spin liquid \cite{Moessner01}. The bipartiteness of the underlying lattice can also influence the type and relevance of low-energy excitations. In Dirac spin liquids, the square lattice formulation is unstable due to relevant monopole excitations, while on the triangular lattice it is stable since these excitations are irrelevant \cite{song_unifying_2019}. Therefore, given the aforementioned differences of dimer and spin models on bipartite and frustrated lattices, it is natural to inquire into the nature of the phase diagram of the unconstrained $\mathbb{Z}_2$ gauge theory on a triangular lattice and whether a spin liquid state is stable.

Our results are summarized in Fig.~\ref{fig:diagram}. We find that the ground state hosts two phases: a vison crystal phase, where a global \(\mathbb{Z}_2\) symmetry is spontaneously broken, and a superfluid phase. While the vison crystal is a deconfined phase, the superfluid is confined and the transition between them is first-order. The article is organized as follows. In the next section, we discuss in detail the symmetries of the model, and classify possible phases. Sec.~\ref{sec:results} presents our numerical results both at $T=0$ and at finite temperature. In Sec.~\ref{sec:conclusions} we conclude and discuss possible generalizations of the model that may allow for various other phases and phase transitions. The article is supplemented by four appendices that discuss a Clausius--Clapeyron relation for our model, the details of the AF-QMC simulations,  the low-energy limit of the \(\pi\)-flux state on the triangular lattice, and a parton mean-field  theory of deconfined phases.

\section{Model, symmetries and method}
\label{sec:modelmethods}

\subsection{The model}
\label{subsec:model}

\begin{figure}[t]
    \centering
    \includegraphics[scale=0.15]{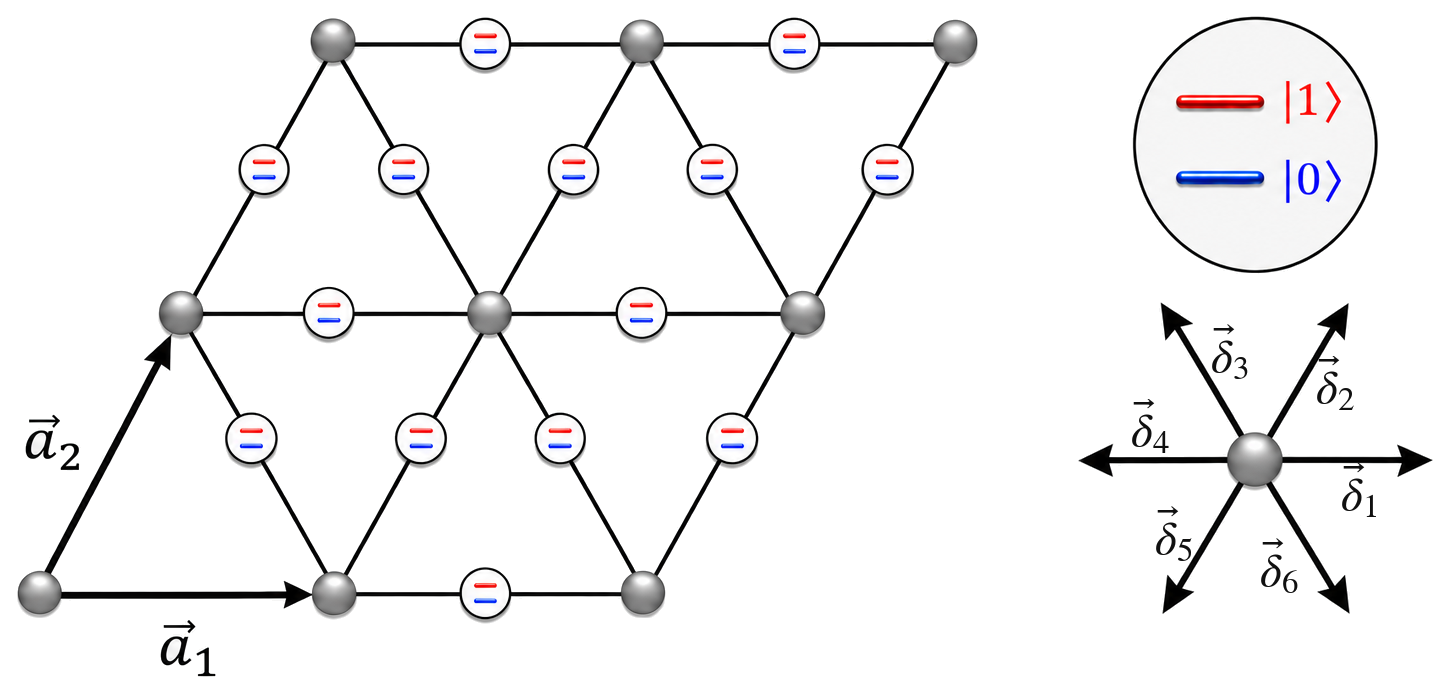}
    \caption{A portion of the triangular lattice with $L=3$. Sites with fermionic degrees of freedom are represented by gray spheres; primitive vectors are denoted by $\mathbf{a}_1 = (1,0)$ and $\mathbf{a}_2 = 1/2(1,\sqrt{3})$. The two-level systems are represented by the circles on each bond $b$. Periodic boundary conditions are assumed.}
    \label{fig:cartoom}
\end{figure}

The Hamiltonian we study reads 
\begin{align}
\label{eq:hamiltonian}
    \hat{H} = \sum_{\ave{\iv,\jv}} \sum_{\s} \hat{Z}_{\ave{\iv,\jv}} \left( \hat{c}_{\iv,\s}^{\dagger} \hat{c}_{\jv,\s}^{\phantom{\dagger}}  + \text{H.c.} \right) -
    h \sum_{\ave{\iv,\jv}}\hat{X}_{\ave{\iv,\jv}}
\end{align}
where the sums run over a triangular lattice, with $\ave{\iv,\jv}$ denoting nearest-neighbor sites. The operator $\hat{c}_{\iv,\s}^{\dagger} (\hat{c}_{\iv,\s}^{\phantom{\dagger}})$ creates (annihilates) a fermion at site $\iv$ with flavor $\s$, \(\{\hat{c}_{\iv,\s}^{\phantom{\dagger}}, \hat{c}_{\jv,\s'}^{\dagger}\} = \delta_{\iv,\jv}\delta_{\s,\s'}\). Each bond $b =  \ave{\iv,\jv}$
 accommodates a two-level system, i.e. Ising fields, whose Hilbert space is spanned by the states $\ket{0}_b$ and $\ket{1}_b$. In this basis, the operators $\hat{Z}_{\ave{\iv,\jv}}$ and $\hat{X}_{\ave{\iv,\jv}}$ correspond to the Pauli matrices $\hat{\s}^{z}$ and $\hat{\s}^{x}$, respectively. The operators satisfy \(\{\hat{Z}_{b}, \hat{X}_{b} \} = 0 \)  and 
 \( \left[ \hat{Z}_{b}, \hat{X}_{b'} \right]  = 0\) if  $b \neq b'$. Here, $h$ tunes the strength of the transverse field, giving dynamics to the \(\mathbb{Z}_2\) gauge field \(\hat{Z}_{\ave{\iv,\jv}}\). We set the hopping amplitude to unity, so that all energy scales, including $h$ and the temperature $T = 1/\beta$, are measured in units of the hopping energy. A schematic representation of our model is shown in Fig.~\ref{fig:cartoom}.

\subsection{Symmetries}
\label{subsec:symmetries}

The model defined in Eq.~\eqref{eq:hamiltonian} leads to a number of symmetries. It has time-reversal, spin SU(2), lattice translation, global particle number  conservation, and \(\mathrm{C}_{6v}\) point group symmetries. The model has a local \(\mathbb Z_2\) gauge symmetry and a global \(\mathbb Z_2\) Ising symmetry which lead to interesting consequences as we detail below.

We start by defining a symmetry that corresponds to the local $\mathbb{Z}_2$ conservation laws: 
\begin{align}
[ \hat{Q}_{\iv}, \hat{H} ] = 0.
\end{align}
For each site $\iv$ the corresponding conserved operator is given by: 
\begin{align}
    \hat{Q}_{\iv} = \left (\hat{X}_{\iv,\iv + \av_1} \hat{X}_{\iv,\iv + \av_2} \hat{X}_{\iv,\iv + \av_2 - \av_1} \right. \times \nonumber\\ 
    \left. \hat{X}_{\iv,\iv - \av_1}\hat{X}_{\iv,\iv - \av_2} \hat{X}_{\iv,\iv + \av_1 - \av_2} \right ) \hat{p}_\iv,
\end{align}
see Fig.~\ref{fig:cartoom}  for a representation of the bonds involved in $\hat{Q}_\iv$. Here $\hat{p}_\iv $ is the parity of fermion number
at site $\iv$:
\begin{align}
    \hat{p}_\iv = \prod_{\s \in \lbrace \uparrow,\downarrow \rbrace} \left( 1-2\hat{n}_{\iv,\s} \right) = \left ( -1 \right)^{\hat{n}_\iv},
\end{align}
where \(\hat n_{\iv} = \sum_\sigma \hat n_{\iv, \s} \) and 
 \(\hat n_{\iv, \sigma} = \hat{c}_{\iv,\s}^{\dagger}\hat{c}_{\iv,\s}^{\phantom{\dagger}}\). The quantity \(\hat Q_\iv\) is a restatement of Gauss's law and it is a conserved quantity. Note that $[\hat{Q}_{\iv},\hat{Q}_{\jv}] = 0$ and $\hat{Q}_{\iv}^{2} = 1$ hold. Its eigenvalues define gauge charge sectors, here named Gauss sectors. Thus the eigenstates of the Hamiltonian \(\ket{\Psi}\) can be written as 
\begin{equation}
    \hat Q_\iv \ket{\Psi\{Q_{\jv}\}} = Q_\iv \ket{\Psi\{Q_{\jv}\}}.
\end{equation} 
The conservation of \(\hat Q_\iv\) implies that the Hamiltonian Eq.~\eqref{eq:hamiltonian} is invariant under a \(\mathbb{Z}_2\) gauge transformation defined by

\begin{equation}
    \hat T_Q = e^{\sum_{\iv} i \pi \theta_\iv \left(  \hat{n}_\iv  + \sum_{n} \frac{1-\hat{X}_{\iv,\iv+ \mathbf{\delta}_n}}{2} \right) } 
    \equiv \prod_{\iv}\hat{Q}_\iv^{\theta_\iv}, 
\end{equation}
where \(\theta_\iv = 0,1\).   In particular, 
\begin{equation}
    \hat{T}_Q^{-1} \hat{c}^{\dagger}_{\iv,\s} \hat{T}_Q =  e^{i \pi \theta_\iv} \hat{c}^{\dagger}_{\iv,\s}
        \quad \text{and}  \quad 
    \hat{T}_Q^{-1} \hat{Z}_{\iv,\jv} \hat{T}_Q =  e^{i \pi \theta_\iv}  e^{i \pi \theta_\jv} \hat{Z}_{\iv,\jv}
\end{equation}
 
\begin{figure}[t]
    \centering
    \includegraphics[scale=0.4]{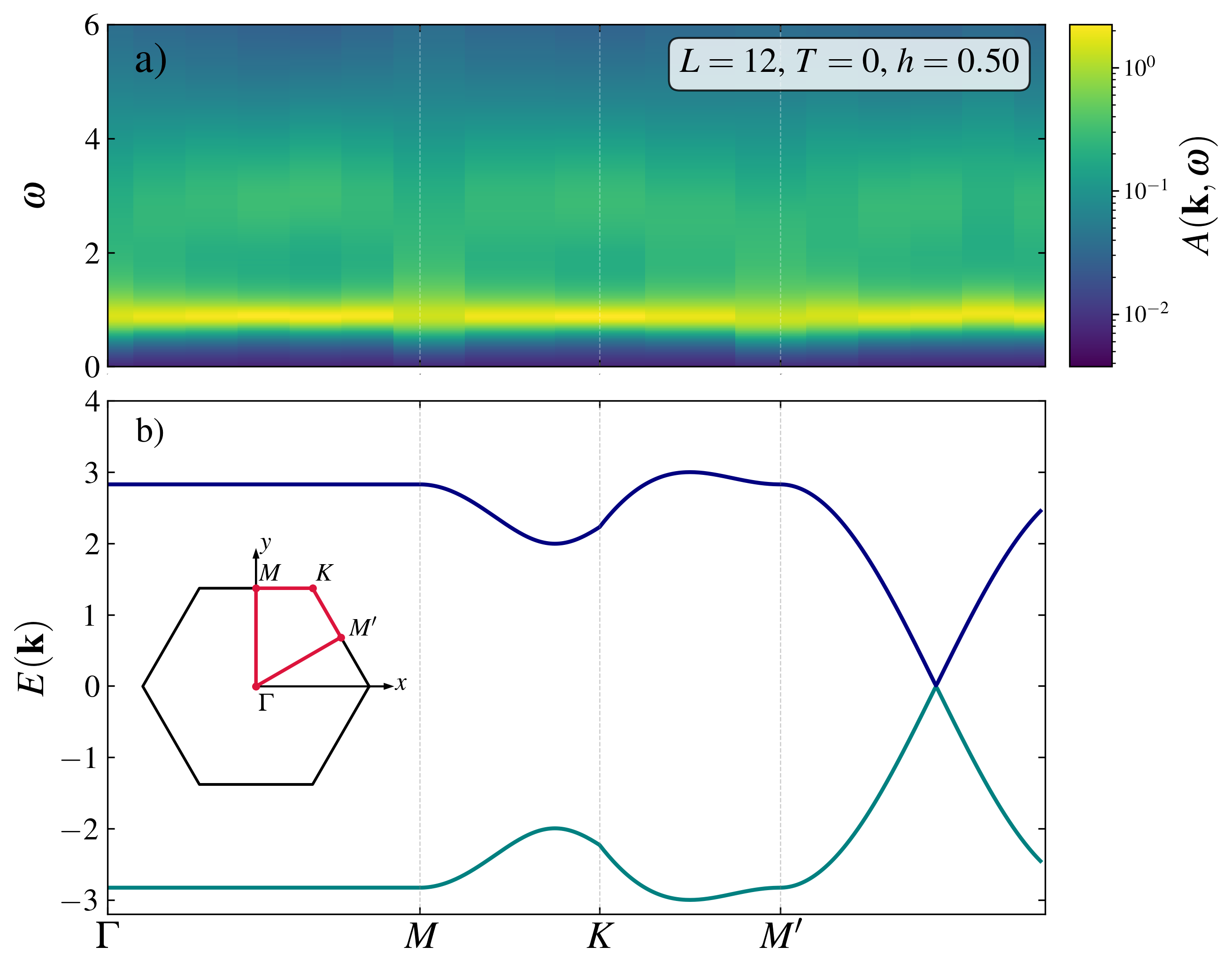}
    \caption{(a) Single-particle spectral function $A(\kv,\omega)$ obtained from projective AF-QMC simulations along the path through the Brillouin zone of the triangular lattice shown in the inset of panel (b). As argued in the text, the local $\mathbb{Z}_2$ symmetry enforces momentum independence. (b) Energy bands of the triangular lattice with a staggered flux pattern ($\pi$-flux). The exact derivation of these energy bands can be found in Appendix~\ref{App:dirac_eq}.}
    \label{fig:green_pi_flux}
\end{figure}

It is important to mention that, despite the extensive number of conserved quantities, these conservation laws do not render the Hamiltonian integrable. Indeed, for an $N = L_x \times L_y$ lattice, the number of fermionic degrees of freedom is $2 N$; additionally, there are $3 N$ degrees of freedom arising from the Ising fields. Therefore, after accounting for the $N$ conserved $Q_\iv$ operators, the remaining number of degrees of freedom of our Hamiltonian is still extensive:
\begin{equation}
   N_{\text{dof}} = 2N + 3N - N = 4N.
\end{equation}
The local $\mathbb{Z}_2$ symmetry has important consequences. Since $[ \hat{H}, \hat{Q}_\iv] = 0$ and $\{ \hat{Q}_\iv, \hat{c}_{\iv,\s}^{\dagger} \} = 0$, it can be shown that 
\begin{equation} \label{eq:z2_green}
    \ave{\hat{c}_{\iv,\s}^{\dagger}(\tau) \hat{c}_{\jv,\s}^{\phantom{\dagger}}(0)} = \delta_{\iv,\jv} \ave{\hat{c}_{\iv,\s}^{\dagger}(\tau) \hat{c}_{\iv,\s}^{\phantom{\dagger}}(0)}
   \end{equation}
and, similarly,
\begin{equation}
    \ave{\hat{Z}_{b}(\tau) \hat{Z}_{b'}(0)} = \delta_{b,b'}  \ave{\hat{Z}_{b}(\tau) \hat{Z}_{b}(0)}.
\end{equation}
This means that a bare fermion $\hat{c}_{\iv,\s}^{\dagger}$ as well as the two-level operator $\hat{Z}_b$ cannot propagate in real space. By contrast, quantities such as
\begin{equation}
    \ave{\hat{c}_{\iv,\s}^{\dagger}  \prod_{b \in\gamma_{\iv\jv}}\hat{Z}_{b} \hat{c}_{\jv,\s}},
\end{equation}
obtained by dressing a fermion with a string of $\hat{Z}$ operators along a $\gamma_{\iv \jv}$ path connecting sites $\iv$ and $\jv$, are allowed by  symmetry. Similarly, closed products of $\hat{Z}$ operators and local expectation values such as $\hat{X}_{\ave{\iv,\jv}}$ are not required to vanish by symmetry. Furthermore, Eq.~\eqref{eq:z2_green} also implies that the single-particle Green's function is $\kv$-independent in momentum space, as depicted in Fig.~\ref{fig:green_pi_flux}(a), and that its imaginary-time dependence measures the gap between Gauss sectors: \(\Delta_{\mathrm G} = E(Q_1, \hdots, -Q_i,  \hdots , Q_N) - E(Q_1, \hdots, Q_N)\).

In addition to the local conservation laws, our model also has a global $\mathbb{Z}_2$ Ising symmetry. Let us define the global transformation

    \begin{align}
    \hat{T} = \prod_{\ave{\iv,\jv}} \hat{X}_{\ave{\iv,\jv}}  \hat{\mathcal P},
\end{align}
where $\hat{\mathcal P}$ is the particle-hole  transformation 
\begin{align}
    \hat{\mathcal P}^{\phantom{-1}} \hat{c}_{\jv,\s}^{\dagger} \hat{\mathcal P}^{-1} =  \hat{c}_{\jv,\s}^{\phantom\dagger}.
\end{align}
With this, it is straightforward to verify that $[\hat{H},\hat{T}] = 0$. This symmetry is closely related to the dynamics of fluxes in our model; let us define the flux through a triangular plaquette as 
\begin{align}
    \hat{F}_{\triangle_\iv} = \prod_{b \in \partial \triangle_\iv} \hat{Z}_b.
\label{eq:op_flux}
\end{align}
Since $[\hat{H},\hat{T}] = 0$, one can show that
\begin{align}
    \ave{ \hat{F}_{\triangle_\iv}} =  \frac{1}{Z}\text{Tr} \Big[ e^{-\beta \hat{H}} \hat{T} \prod_{b \in \partial \triangle_\iv} \hat{Z}_b \hat{T}^{-1} \Big] = (-1)^{N_e}\ave{ \hat{F}_{\triangle_\iv}},
\end{align}
where $N_e$ is the number of edges of the plaquette. Therefore, for lattices whose elementary plaquettes contain an odd number of edges, such as the triangular lattice, the global $\mathbb{Z}_2$ symmetry enforces $ \ave{ \hat{F}_{\triangle_\iv}} = 0$.   Each unit cell accounts for an upward- and a downward-pointing triangle and  similar  statements hold for $\hat{F}_{\bigtriangledown_\iv}$ associated with downward-pointing triangles.

It is worth emphasizing that, although the Hamiltonian in Eq.~\eqref{eq:hamiltonian} closely resembles a $\mathbb{Z}_2$ gauge theory coupled to matter fields \cite{Kogut1975,Fradkin1979,Senthil2000}, it is not a gauge theory in the conventional sense \cite{Assaad2016}. The crucial distinction is that we do not impose Gauss's law as a local constraint. In a \textit{constrained} $\mathbb{Z}_2$ gauge theory, $\hat X_{\iv,\iv + \mathbf{\hat{n}}}$ plays the role of the $\mathbb{Z}_2$ electric field along direction $\mathbf{\hat{n}}$ at site $\iv$. Gauss's law relates its local divergence to the matter charge. Satisfying Gauss's law corresponds to imposing a local constraint on the products of $\hat{X}$ operators around each site, or, equivalently, to projecting the Hilbert space onto a fixed eigensector of the conserved operators $\hat{Q}_\iv$. In our model, no such projection is made, such that we study an \textit{unconstrained} $\mathbb{Z}_2$ gauge theory. The operators $\hat{Q}_\iv$ are conserved quantities, but their eigenvalues are not fixed externally by a gauge constraint.  

We conjecture that Gauss's law will be dynamically imposed \cite{xu_monte_2019,inacio_unconstrained_2026}. The variables $\hat{Q}_\iv$ are classical variables:  they commute with the Hamiltonian and hence possess no quantum dynamics. In the renormalization group sense, a symmetry-allowed interaction between these classical variables will be generated and will take the generic form:
\begin{equation} \label{eq:model_gauss_ops}
    \hat{H}_Q  = h_Q \sum_{\iv} \hat{Q}_\iv   + \sum_{\iv,\jv} J_Q(\iv-\jv) \hat{Q}_\iv \hat{Q}_\jv + \cdots.
\end{equation} 
In the above, we have assumed translation invariance and omitted three-body and higher-order terms. Since no symmetry forces $\langle \hat{Q}_\iv \rangle$ to vanish, a field $h_Q$ will be generated. Provided that the above Hamiltonian has a unique ground state, Gauss's law will be dynamically imposed. Below this dynamically generated energy scale the model will be equivalent to a constrained $\mathbb{Z}_2$ gauge theory. 
To probe the  validity of the above  effective Hamiltonian, we have computed $\ave{\hat{Q}_\iv}$ to  obtain $1.01 \pm 0.01$ at $h = 4$ within a ground state calculation  \footnote{Note that the  trial wave function used in the calculation was not an eigenstate of $\hat{Q}_{\iv}$}. This shows that at this coupling strength $\hat{H}_Q$ corresponds to a ferromagnetic Ising model subject to a uniform magnetic field.

Finally, we  note that $\hat{T}$ commutes with $\hat{Q}_{\iv}$ for our SU(2) spin-symmetric model.  Therefore, even  when the global $\hat{T}$ Ising symmetry is spontaneously broken the effective Ising model for Gauss's law, Eq.~\eqref{eq:model_gauss_ops}, is still valid.

\subsection{The method}
\label{subsec:dqmc}

We investigate the physical properties of our model, Eq.~\eqref{eq:hamiltonian}, using the auxiliary-field quantum Monte Carlo method (AF-QMC)  \cite{Blankenbecler1981,Hirsch1983,Hirsch1985,Kawashima2002,rrds2003,assaad_alf_2025}, as implemented in the ALF package \cite{assaad_alf_2025}. Ground state properties are obtained using projective AF-QMC, while finite-temperature properties are studied using the finite-temperature formulation. 
Both approaches are free of the negative sign problem for our model. Further details on the method and its implementation can be found in Appendix \ref{App:qmc_details}. 

Here, we investigate the occurrence of a superfluid (SF) phase through the analysis of the pairing structure factor,
\begin{align}
    S_{\mathrm{SF}} (\qv) = \frac{1}{N}\sum_{\iv,\jv}e^{i\qv \cdot (\iv - \jv)}  \ave{\hat{\Delta}_\iv^{\dagger} \hat{\Delta}_\jv^{\phantom{{\dagger}}} + \hat{\Delta}_\jv^{\dagger}\hat{\Delta}_\iv^{\phantom{{\dagger}}}},
\end{align}
where $\hat{\Delta}_\iv^{\dagger} = \hat{c}_{\iv,\uparrow}^{\dagger}\hat{c}_{\iv, \downarrow}^{\dagger}$ is the $s$-wave pairing operator. The ordering wave vector is $\mathbf{\Gamma}=(0, 0)$ which corresponds to the $s$-wave state. Similarly, we examine the emergence of a vison crystal phase by analyzing the connected flux structure factor:
\begin{align}
    S_{\mathrm{F}} (\qv) = \frac{1}{N}\sum_{\alpha} \sum_{\iv, \jv} e^{i\qv \cdot (\iv - \jv)} \left( \ave{\hat{F}_{\iv}^{\alpha} \hat{F}_{\jv}^{\alpha}} - \ave{\hat{F}_{\iv}^{\alpha}}\ave{\hat{F}_{\jv}^{\alpha}} \right),
\end{align}
where $\hat F_{\iv}^{\alpha}$ is the flux operator defined in Eq.~\eqref{eq:op_flux}, and $\alpha \in \lbrace \triangle, \bigtriangledown \rbrace$ denotes the two triangles associated with each unit cell.  With this convention,  intra-unit-cell flux  patterns show up as Bragg peaks  at the
 $\mathrm{\mathbf{\Gamma}}=(0, 0)$  point in the  Brillouin zone, and  hence do not break translation symmetries.

Phase boundaries are determined by the correlation ratio \cite{Kaul2015,Liu2018,Sato2018,Otsuka2020,Kennedy2025}
\begin{align}
    R_{\mathrm O}(L,h,T) = 1 - \frac{S_{\mathrm O}(\qv_{\mathrm O} + \delta\qv)}{S_{\mathrm O} (\qv_{\mathrm O})}
\label{eq:rc}
\end{align}
where $\qv_{\mathrm O}$ is the ordering wave vector of a phase associated with $S_{\mathrm O}$, $\mathrm O$ denotes ``SF'' or ``F'', and $|\delta\qv| \sim 2\pi/L$ is the shortest wave vector away from $\qv_{\mathrm O}$. The correlation ratio is a renormalization group invariant quantity that approaches one in the ordered phase and zero in the disordered phase. At the critical point, we expect \(R_{\mathrm O}(L,h,T) = f_{\mathrm O}(L^z/\beta, (h - h_c)L^{1/\nu}, L^{-\omega})\), where \(z\) is the dynamical exponent, \(h_c\) is the value of the critical coupling, \(\nu\) is the correlation-length exponent  and \(\omega\) is the leading correction-to-scaling exponent. Since we have Lorentz symmetry, \(z = 1\), and we will ignore drift effects set by irrelevant operators. Therefore, curves of $R_{\mathrm O}(h,T)$ for different $L$ cross at $h_c$ ($T_c$), thus providing estimates for the critical points that converge to the exact value as $L\rightarrow \infty$.

\section{Limiting cases}
\label{sec:limits}

Here we will discuss possible phases at  weak and  strong coupling.

\subsection{Vanishing transverse field limit: \(h = 0\)}

At \(h = 0\), the link variables \(\hat{Z}_{\langle \iv, \jv \rangle}\) become classical:
\begin{equation}
    \hat H_{h=0} = \sum_{\langle \iv, \jv \rangle, \s} Z_{\langle \iv, \jv \rangle} (\hat{c}_{\iv,\s}^{\dagger} \hat{c}_{\jv,\s}^{\phantom{\dagger}}  + \text{H.c.} ).
\end{equation}
The ground state is found by minimizing the energy as a function of the link variables, \(Z_{\langle \iv, \jv \rangle} = \pm 1\). It turns out that at half filling, the configuration with minimal ground-state energy is the
\(\pi\)-flux pattern considered in Ref.~\cite{Otsuka18}: $F_{i_{\bigtriangledown}} = - F_{i_{\triangle}} $.  We note that Lieb's theorem \cite{lieb_flux_1994} does not apply to this  frustrated lattice, and that the above statement stems from numerical calculations on small lattices.  For a  fixed choice of link  variables that produces the aforementioned  flux pattern, the \(c\)-fermions acquire a Dirac dispersion  at half filling, see Fig.~\ref{fig:green_pi_flux}(b).

At $h=0$ the  ground state is  degenerate. This stems from the fact that  the  Hamiltonian commutes with the link variables such  that  if $|\Psi(\{Q_{\iv}\})\rangle$ is an eigenstate, then so is $ \hat{Z}_{ \langle \iv_1, \iv_2 \rangle}  \hat{Z}_{ \langle \iv_2, \iv_3 \rangle} \cdots \hat{Z}_{ \langle \iv_{n-1}, \iv_n \rangle} |\Psi(\{Q_{\iv}\})\rangle$, with the same energy, but in the Gauss sector in which the charges $Q_{\iv_1}$ and $Q_{\iv_n}$ at the endpoints of the string are reversed. Indeed, $\hat{Z}_{\langle \iv, \jv \rangle}$ anticommutes with $\hat{Q}_{\iv}$ and $\hat{Q}_{\jv}$, and every  site not at the boundary of the string is visited twice. In the thermal average over this degenerate manifold, all gauge-dependent quantities will vanish, $\langle \hat{Z}_{\langle \iv, \jv \rangle} \rangle = 0$  
and the single-particle Green's function remains dispersionless, see  Eq.~\eqref{eq:z2_green}.
This is an example of  Elitzur's theorem: any local operator that is not invariant under a local symmetry has a vanishing expectation value \cite{Elitzur75,Batista05}.

\subsection{Weak coupling limit, $h \ll 1$: Parton construction}

\label{subsec:weak-coupling}

A mean-field picture of our model is accessible within a parton construction \cite{Ruegg10,Nandkishore12} that reads:  
\begin{equation}
     \hat{c}_{\iv,\s} = \hat{f}_{\iv,\s} \hat{\tau}_{\iv}^{z}   \quad \text{with} \quad  
     \hat{g}_{\iv} \equiv \hat{\tau}_{\iv}^{x} (-1)^{\hat{n}_{\iv}}  = 1.
\end{equation}
Here, 
$\hat{n}_{\iv} = \sum_{\s}\hat{f}_{\iv,\s}^{\dagger}\hat{f}_{\iv,\s}^{\phantom{\dagger}} = \sum_{\s}\hat{c}_{\iv,\s}^{\dagger}\hat{c}_{\iv,\s}^{\phantom{\dagger}}$. 
The constraint ensures that the Hilbert space of the $c$-fermions and that of the matter field $\hat{\tau}$ and the $f$-fermions have the same dimension.  It leads to the identification:
\begin{align}
         | \uparrow \rangle   & \leftrightarrow  | - ,  \uparrow \rangle    \nonumber \\
         | \downarrow \rangle   & \leftrightarrow  | - ,  \downarrow \rangle  \nonumber \\
         | 0           \rangle   & \leftrightarrow  | + ,  0 \rangle    \nonumber \\
         | \uparrow\downarrow \rangle   & \leftrightarrow  | + ,  \uparrow\downarrow \rangle  
\end{align}
Since $\hat{\tau}_{\iv}^{x} (-1)^{\hat{n}_{\iv}}  = 1$, the conserved charge $\hat{Q}_{\iv}$ now reads:
\begin{equation}
    \hat{Q}_{\iv} = \hat{\tau}_{\iv}^{x}  \prod_{\delta} \hat{X}_{\iv,\iv + \mathbf n_\delta},
\end{equation}
where $\delta$ runs over the six nearest-neighbor vectors. Thereby, the $f$-fermion carries no $\hat{Q}_{\iv}$ charge, 
and within this parton construction, the Hamiltonian reads:
\begin{equation}
 \label{eq:hamiltonian_f}
    \hat{H} = \sum_{\ave{\iv,\jv}} \sum_{\s} \left( \hat{f}_{\iv,\s}^{\dagger}  
    \hat{\tau}_{\iv}^{z}\hat{Z}_{\ave{\iv,\jv}} \hat{\tau}_{\jv}^{z} 
    \hat{f}_{\jv,\s}^{\phantom{\dagger}}  + \text{H.c.} \right) -
    h \sum_{\ave{\iv,\jv}}\hat{X}_{\ave{\iv,\jv}}. 
\end{equation}
The above is exact. In particular, the $f$-fermions are subject to an emergent $\mathbb{Z}_2$ gauge field, $\hat{\tau}_{\iv}^{z}\hat{Z}_{\ave{\iv,\jv}} \hat{\tau}_{\jv}^{z}$, and the constraint $\hat{g}_{\iv} = \hat{\tau}_{\iv}^{x} (-1)^{\hat{n}_{\iv}}  = 1$ acts as an emergent Gauss law for the emergent gauge charge. Moreover, one can check that $\hat{g}_{\iv}$ is locally conserved such that the $f$-fermions cannot propagate freely in space and time. In contrast, $\hat{Q}_{\iv}$ is conserved but not constrained, such that the $c$-fermion propagator is local in space but not in imaginary time. In Appendix~\ref{App:Mean-field} we show that if we assume that $\hat{Q}_{\iv}=1$, then the parton and original models are identical.

\paragraph{Deconfined phase}
The above formulation allows one to carry out a mean-field analysis of the deconfined phase of the emergent $\mathbb{Z}_2$ gauge theory.
When the emergent gauge field is deconfined, a pair of $f$-fermions with attached emergent $\mathbb{Z}_2$ strings can propagate freely. This physics can be captured within a mean-field theory in which the constraint
$\hat{g}_{\iv} = 1$ is enforced on average. Under this approximation, the $f$-fermions move in a background non-fluctuating gauge field, and the fermionic part of the mean-field Hamiltonian (see Appendix~\ref{App:Mean-field}) takes the form: 
\begin{equation}
       \hat{H}_{\text{MF}}^{F} = \sum_{\ave{\iv,\jv}} \sum_{\s} \left( \hat{f}_{\iv,\s}^{\dagger}  
       e_{\ave{\iv,\jv}} 
       \hat{f}_{\jv,\s}^{\phantom{\dagger}}  + \text{H.c.} \right),
\end{equation}  
with $e_{\ave{\iv,\jv}} = \langle \hat{\tau}_{\iv}^{z}\hat{Z}_{\ave{\iv,\jv}} \hat{\tau}_{\jv}^{z} \rangle$. In this mean-field approach, the $c$-fermions are localized in space since $\hat{Q}_{\iv}$ remains a conserved quantity. On the other hand, the $f$-fermions can propagate in space and time. This is very similar to the orthogonal metals discussed in Refs.~\cite{Nandkishore12,Hohenadler2018,Hohenadler2019}. 

\paragraph{Vison crystal}
For small values of $h$ we expect the $e_{\ave{\iv,\jv}}$ to account for a $\pi$-flux pattern, i.e., $F_{i_{\bigtriangledown}} = - F_{i_{\triangle}}$ leading to Dirac fermions (see Appendix~\ref{App:dirac_eq}). We call this phase a vison crystal, since it breaks a point group symmetry of the original lattice. In particular, mirror symmetries along the lattice vectors that map upward to downward triangles are broken. At the mean-field level, where one chooses a particular gauge, translation symmetry is broken in the sense that the unit cell consists of two sites. The $\pi$-flux per unit cell implies that translation operators along the unit vectors, $\hat{T}_{\mathbf a_{1/2}}$ anticommute: $\hat{T}_{\mathbf a_1}\hat{T}_{\mathbf a_2} = -\hat{T}_{\mathbf a_2}\hat{T}_{\mathbf a_1}$. 

Our numerical simulations go beyond the mean-field approximation, and take into account the $\hat{g}_{\iv}=1$ constraint exactly. The translation of the $f$-fermion operator is then defined up to a local gauge transformation. In other words, only gauge invariant notions remain physically meaningful. In the mean-field theory the choice of the two-orbital unit cell is a gauge choice  that becomes meaningless in an exact calculation. Thereby, all physical quantities remain invariant under translations along the lattice vectors.  

In the deconfined phase of the emergent gauge field, the $f$-fermions with appropriate $\mathbb{Z}_2$ string attachments can propagate freely. Assume that they generate a Dirac spectrum. Then Luttinger's theorem \cite{Oshikawa00a} that applies to the $f$-fermions will be violated. This is the hallmark of the deconfined phase.

\subsection{Strong transverse field limit: \(h \to \infty\)}
\label{subsec:strong_h}
In the limit of $h\rightarrow \infty$, our model is equivalent to  a Su-Schrieffer-Heeger (SSH) model coupled to an Einstein phonon mode of frequency $h$ \cite{Su1980,Assaad2016,Goetz2024}. In this limit, we write the effective Hamiltonian as
\begin{align}
    \hat{H}_{\infty} = - \frac{1}{2h}\sum_{\ave{\iv,\jv}}\hat{K}_{\ave{\iv,\jv}}^2
\label{eq:ham_strong_h}
\end{align}
where
\begin{align}
\hat{K}_{\ave{\iv,\jv}} = \sum_{\s \in \lbrace \uparrow,\downarrow \rbrace} \left( \hat{c}_{\iv,\s}^{\dagger} \hat{c}_{\jv,\s}^{\phantom{\dagger}}  + \hat{c}_{\jv,\s}^{\dagger} \hat{c}_{\iv,\s}^{\phantom{\dagger}} \right).
\end{align}

It is instructive to understand the nature of the phases and the role of the non-bipartite nature of the triangular lattice. For a general lattice, the strong-coupling effective interaction can be decomposed into spin and ``Anderson pseudospin'' sectors \cite{Anderson1958},
\begin{align}
    \hat{H}_{\infty} = \frac{2}{h}\sum_{\ave{\iv,\jv}} \left[\mathbf{\hat{S}_{\iv}} \cdot \mathbf{\hat{S}_{\jv}} + \hat{\eta}_{\iv}^{z}\hat{\eta}_{\jv}^{z} -\frac{1}{2} (\hat{\eta}_{\iv}^{+}\hat{\eta}_{\jv}^{-} + \hat{\eta}_{\iv}^{-}\hat{\eta}_{\jv}^{+})\right]
\label{eq:ham_strong_h_spin}
\end{align}
where, 
\begin{align*}
    \hat{\eta}_{\iv}^{z} = 1/2(1 - \hat{n}_\iv), \quad  \hat{\eta}_{\iv}^{+} = \hat{c}_{\iv, \uparrow}^{\dagger}\hat{c}_{\iv, \downarrow}^{\dagger}, \quad  \hat{\eta}_{\iv}^{-} = (\hat{\eta}_{\iv}^{+})^{\dagger}.
\end{align*}
The pseudospin part does not, in general, take the isotropic form $\hat{\boldsymbol{\eta}}_\iv\cdot\hat{\boldsymbol{\eta}}_\jv$. Instead, one obtains an anisotropic interaction of the form of Eq.~\eqref{eq:ham_strong_h_spin}, which is antiferromagnetic in the charge-density-wave component but has the opposite sign in the transverse pairing components. On a bipartite lattice, such as the square lattice \cite{Assaad2016}, this relative sign can be removed by a staggered unitary transformation, $\hat{\eta}_i^\pm\rightarrow \epsilon_i\hat{\eta}_i^\pm$, with $\epsilon_i=\pm1$ on the two sublattices and $\epsilon_i\epsilon_j=-1$ for every nearest-neighbor bond. This maps the pseudospin interaction to the isotropic form $\hat{\boldsymbol{\eta}}_\iv\cdot\hat{\boldsymbol{\eta}}_\jv$, thereby making explicit the enlarged symmetry relating antiferromagnetism (AFM), charge density wave (CDW) order, and $s$-wave pairing (SF). On a non-bipartite lattice, such as the triangular lattice, there is no such transformation. Consequently, the exact pseudospin SU(2) structure is broken, and the degeneracy between AFM, CDW and SF states is no longer symmetry protected.

In the  strong  coupling limit,  we have checked  explicitly  with QMC simulations that \(\langle \hat{Q}_{\iv} \rangle = 1\) for all sites \(\iv\) in the low-temperature regime. Hence our model is promoted to a constrained $\mathbb{Z}_2$ gauge  theory.   Any  ordered  phase, $O$, be it in the particle-particle  or particle-hole channels  will not carry any  gauge  charge  such that the  gauge sector can be treated independently.  It  can be deconfined or confined thereby defining  $O^{\star}$ or $O$ phases respectively.

\section{Results}
\label{sec:results}

In this section we present the results of our study. We will start with a summary of the phase diagram as shown in Fig.~\ref{fig:diagram}. Then we will discuss the ground state properties along with their dynamical properties. In the end, we study the temperature dependence of our ground state analysis.

\subsection{Phase diagram}

We summarize our results in the $T \times h$ phase diagram shown in Fig.~\ref{fig:diagram}. In the ground state, we find two distinct phases. For small transverse fields, the system realizes a vison crystal phase, which hosts deconfined fermions carrying a local emergent $\mathbb{Z}_2$ charge. That is, the single-particle spectrum of the \(c\)-fermions is dispersionless, while gauge-invariant quantities, such as spin correlations, retain features of Dirac physics. Upon increasing $h$, quantum fluctuations suppress the flux order and favor an $s$-wave superfluid phase. We also determined, through the analysis of the correlation ratios, that these phases are separated by a first-order quantum phase transition at $h_c \simeq 0.15$.

Thermal fluctuations suppress both orders at high enough temperatures. However, the two phases melt with different critical behaviors due to the distinct global symmetries they break. The superfluid phase realizes quasi-long-range order at finite temperatures and undergoes a Kosterlitz--Thouless (KT) transition. In contrast, the vison crystal phase melts through a continuous transition belonging to the two-dimensional Ising universality class.  As shown in Fig.~\ref{fig:diagram}, the corresponding thermal phase boundaries exhibit different dependences on the transverse field. In the following subsections, we present and discuss the numerical results supporting the phase diagram. We start with the ground-state properties and subsequently discuss the excitation spectrum and then the finite-temperature transitions.

\subsection{Ground state}
\label{subsec:GS}

\begin{figure}[t]
    \centering
    \includegraphics[scale=0.22]{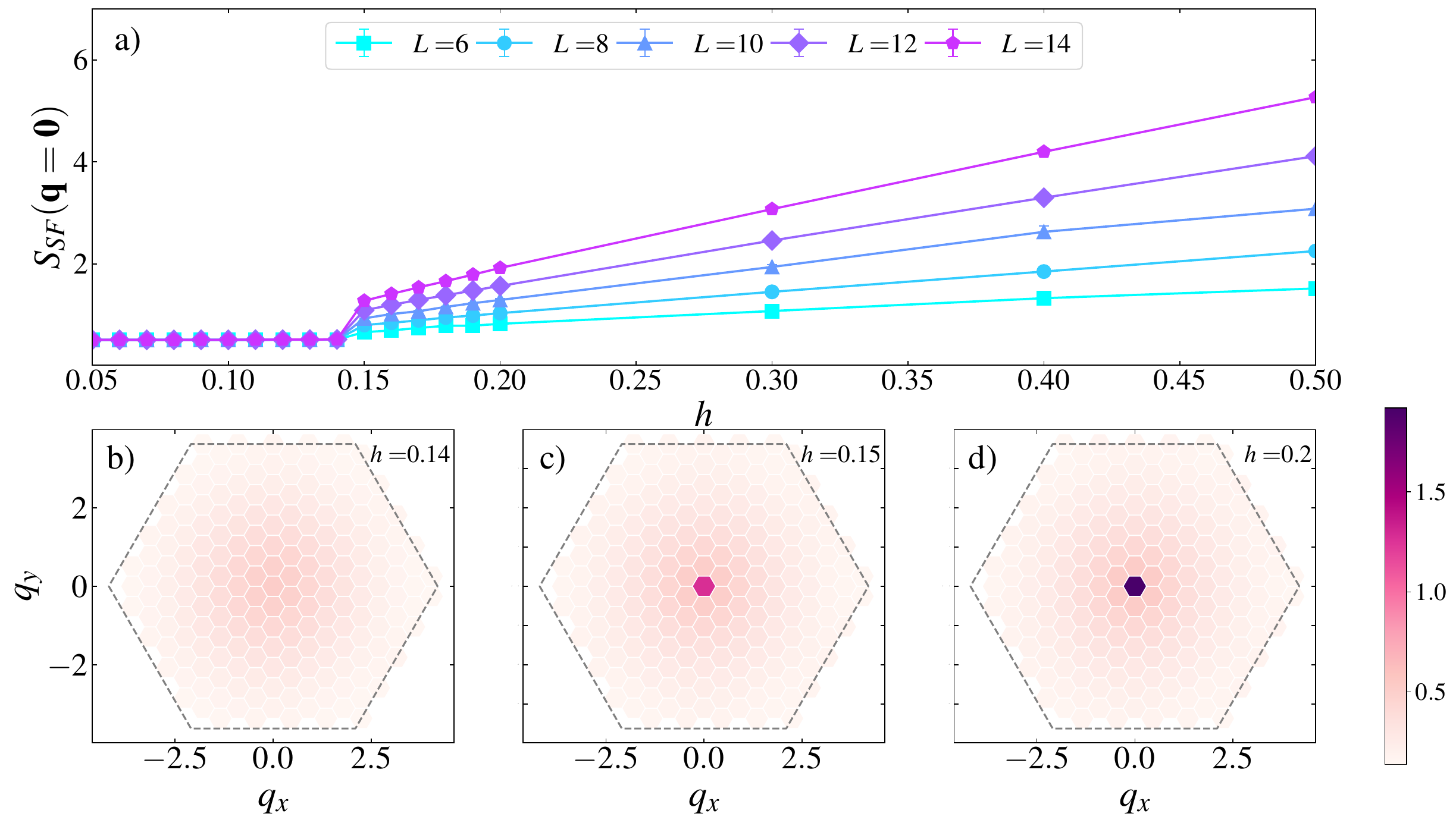}
    \caption{Pairing structure factor. (a) Pairing structure factor at $\qv=(0,0)$ as a function of the transverse field $h$ for different system sizes. (b)--(d) Momentum-resolved pairing structure factor computed for all momenta of a $14\times 14$ lattice for fixed $h$ values. The sudden disappearance of the peak at $\qv=(0,0)$ for $h\lesssim0.15$ signals a first-order phase transition.}
    \label{fig:S_sc}
\end{figure}

\begin{figure}[h]
    \centering
    \includegraphics[scale=0.38]{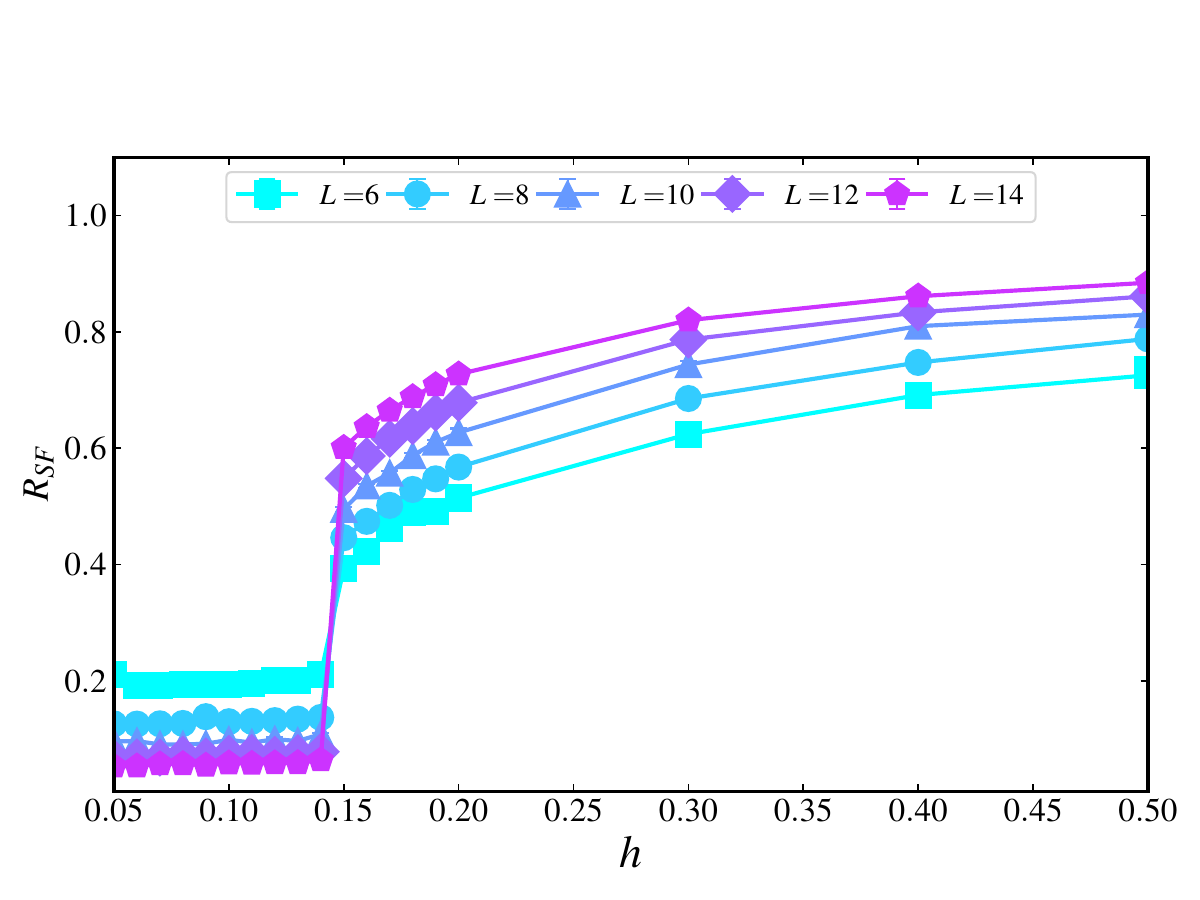}
    \caption{Correlation ratio for the superfluid phase, ordering vector $\qv = (0,0)$, calculated for lattices up to $L=14$. The sharp decrease of $R_{\mathrm{SF}}$ to values close to zero for $h\lesssim0.15$ confirms that the superfluid order is suppressed through a first-order phase transition.}
    \label{fig:R_sc}
\end{figure}

At zero temperature, thermal fluctuations are absent and the ground-state properties are controlled by quantum fluctuations induced by the transverse field $h$. For a bipartite lattice, such as the square lattice, a sufficiently large $h$ drives the system into a threefold-degenerate AFM/SF/CDW state \cite{Assaad2016}. By contrast, on a non-bipartite lattice, this degeneracy is lifted, and the SF state becomes energetically favored. We therefore begin our analysis by examining superfluid correlations through the pairing structure factor. As shown in Fig.~\ref{fig:S_sc}(d), for sufficiently large $h$ the momentum-resolved structure factor develops a pronounced peak at $\qv=(0,0)$, consistent with $s$-wave superfluid order. Moreover, Fig.~\ref{fig:S_sc}(a) shows that this peak grows with system size for $h \gtrsim 0.15$, providing further evidence for long-range SF order. Finally, we compute the SF correlation ratio, which offers a more stringent diagnostic. As shown in Fig.~\ref{fig:R_sc}, in this range of $h$, $R_{\mathrm{SF}}$ takes values close to unity. These results confirm that the system stabilizes an $s$-wave superfluid phase for $h\gtrsim0.15$.

Upon decreasing the transverse field, Fig.~\ref{fig:S_sc} shows that the superfluid order is suppressed. In particular, panels (b) and (c), together with the finite-size analysis in panel (a), show that the structure-factor peak abruptly disappears for $h \lesssim 0.15$. This behavior is further confirmed by the SF correlation ratio, which drops sharply to values close to zero for $h \lesssim 0.15$, as shown in Fig.~\ref{fig:R_sc}. At a continuous transition, the correlation ratio is expected to follow a finite-size scaling form, with curves for different system sizes crossing near the critical point. In contrast, $R_{\mathrm{SF}}$ develops an increasingly abrupt drop near $h\simeq0.15$, without the crossing behavior characteristic of a continuous critical point. This discontinuous behavior signals a first-order transition out of the superfluid phase.

\begin{figure}[h]
    \centering
    \includegraphics[scale=0.22]{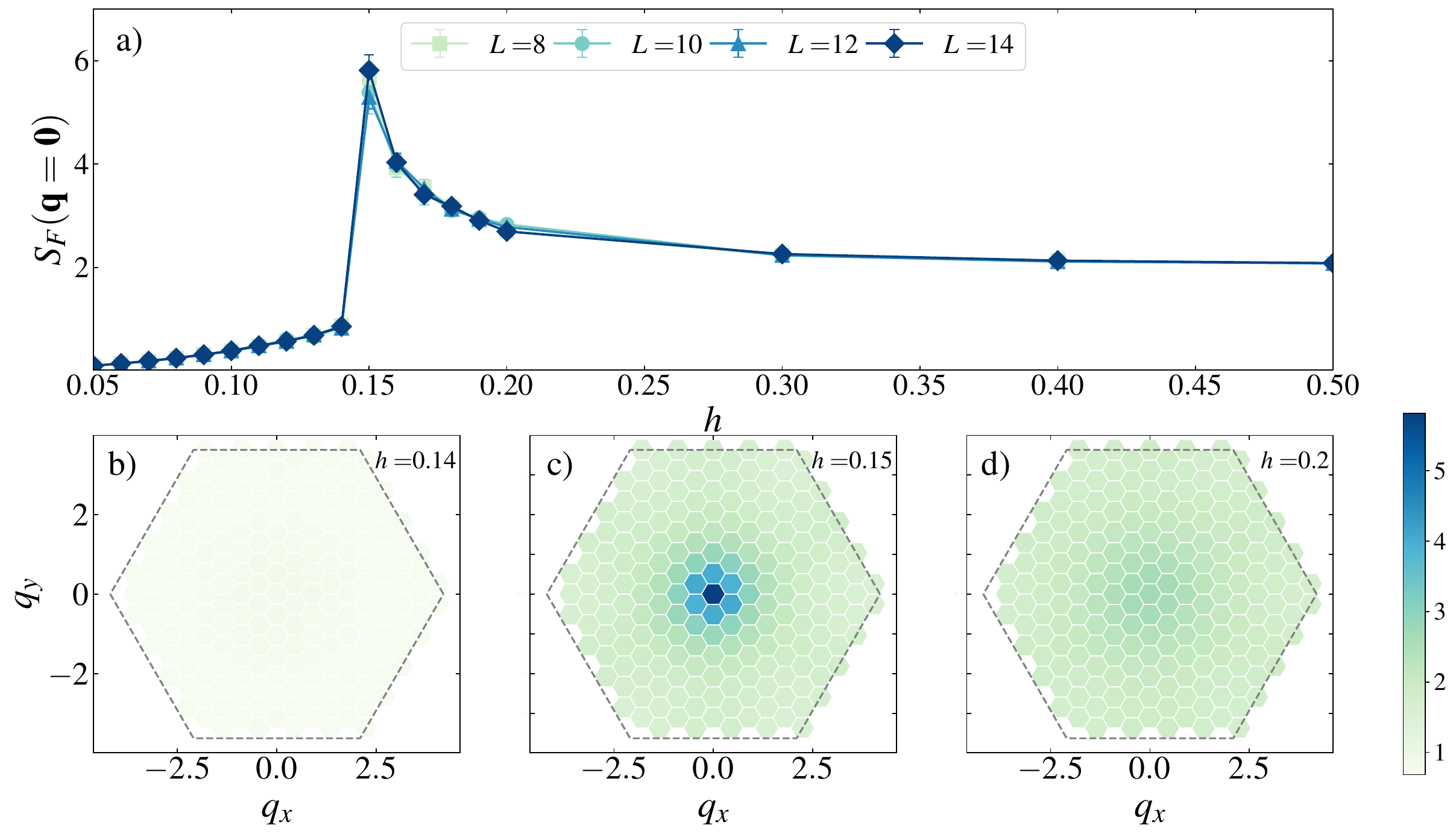}
    \caption{Flux structure factor. (a) Flux structure factor at $\mathbf{q}=(0,0)$ as a function of the transverse field $h$ for different system sizes. (b)--(d) Momentum-resolved flux structure factor computed for all momenta of a $14\times14$ lattice for fixed values of $h$. The enhancement of flux correlations around $h\simeq0.15$ marks the onset of the vison crystal phase, which replaces the superfluid state. The peak at $\mathbf{q}=(0,0)$ indicates a staggered $\pi$-flux pattern.}
    \label{fig:S_F}
\end{figure}

The suppression of superfluidity is accompanied by the emergence of flux order. Figure~\ref{fig:S_F} shows that the flux structure factor is strongly enhanced around $h \simeq 0.15$. In the momentum-resolved data, the dominant peak appears at $\mathbf{q}=(0,0)$. This signals the emergence of long-range flux order. Interpreting the flux excitations as visons, the resulting ordered state can therefore be identified as a vison crystal phase. In particular, as mentioned previously, it breaks a point group symmetry corresponding to  mirror reflections along the lattice  vectors. It is important to note that, since $h$ induces quantum fluctuations, the Ising degrees of freedom become nearly frozen in the weak $h$ regime. This freezing of quantum fluctuations is evidenced in Figs.~\ref{fig:S_F}(a)--(b): once the system no longer fluctuates among different flux configurations, the flux structure factor vanishes, with $S_F$ tending to zero both at $\qv=(0,0)$ and throughout the rest of the Brillouin zone. 
We note that in the  vicinity of the transition, $S_F$ has  little system-size dependence, (see Fig.~\ref{fig:S_F}(a)), indicating that  the  fluctuations of the flux degrees of freedom are bounded. This is a characteristic of a first-order transition.
Therefore, the phase that replaces the superfluid is a vison crystal phase.

\begin{figure}[t]
    \centering
    \includegraphics[scale=0.38]{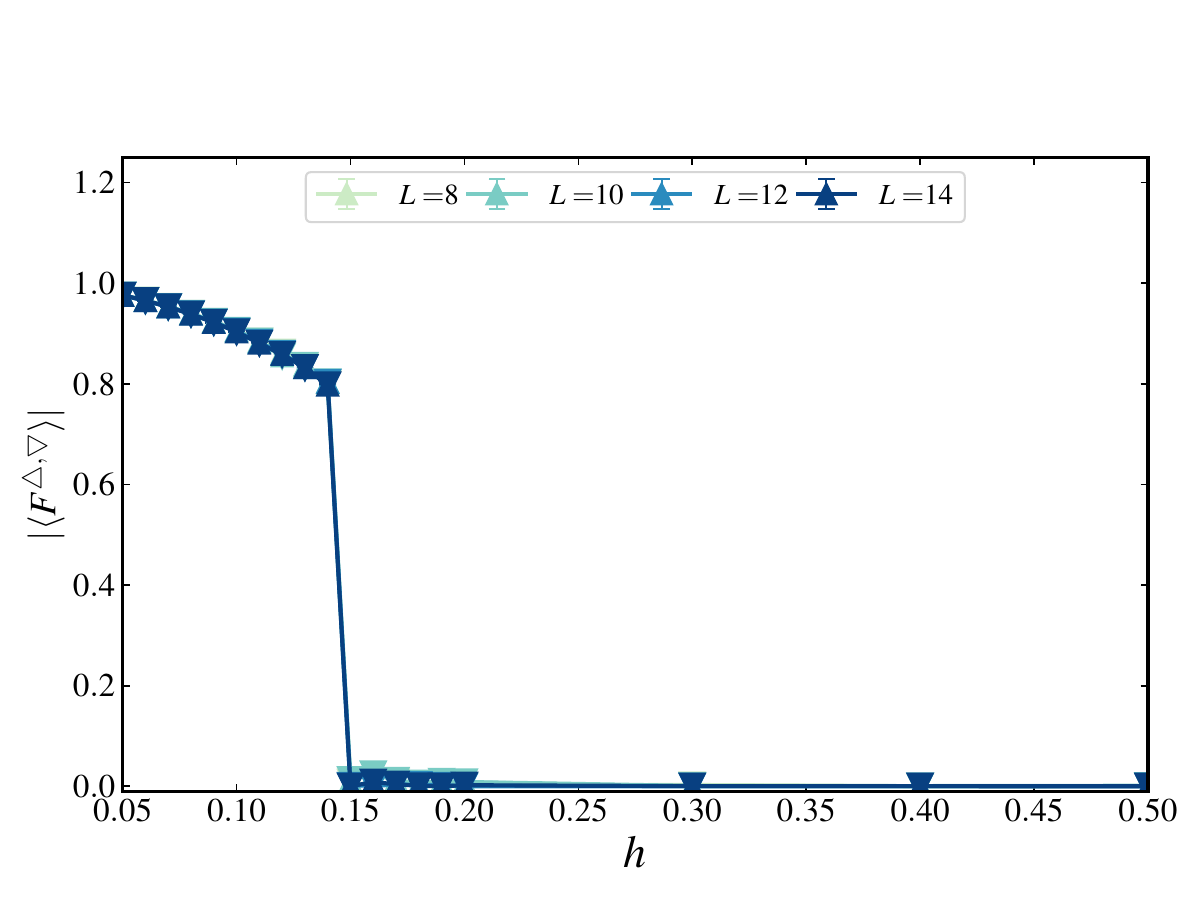}
    \caption{Absolute value of the average flux of the sublattices up (upward triangles) and down (downward triangles). The behavior of this quantity makes it explicit that the global $\mathbb{Z}_2$ symmetry is suddenly broken for $h\lesssim 0.15$.}
    \label{fig:avg_flux}
\end{figure}

To further investigate the symmetry of the vison crystal phase,   we   consider 
the absolute value of the flux on the upward- and downward-pointing triangles separately:  $|\langle \hat{F}^{\triangle} \rangle| =  
\left| \frac{1}{N}\sum_{\iv} \langle  \hat{F}_{\triangle_{\iv}} \rangle \right| $ and similarly for $\hat F^{\bigtriangledown}$. 
By  symmetry,  this  quantity should vanish since 
$\hat{F}_{\triangle_{\iv}}$ is odd under the global Ising symmetry.  However, in the symmetry-broken phase, our single-spin-flip simulations become frozen, and  as shown in Fig.~\ref{fig:avg_flux}, the absolute value of the flux becomes finite. This is very similar to the behavior of the total magnetization in the ordered phase of an Ising ferromagnet when using single-spin-flip updating schemes.

\subsection{Spectral properties }
\label{subsec:dynamics}

The static observables discussed above identify two ground-state phases with distinct broken symmetries, separated by a first-order transition near $h\simeq 0.15$. We next characterize the excitation spectra on both sides of the transition. All the data for the spectral functions shown here were obtained by applying the stochastic analytic continuation (SAC) method \cite{Sandvik1998,beach2004,Shao2023}, as implemented in the ALF package \cite{assaad_alf_2025}, to the imaginary-time correlation functions measured with AF-QMC simulations. For the single-particle spectral function $A(\kv,\omega)$, we use the SAC method to solve the integral equation 
\begin{align}
    G(\kv,\tau) = \int_{-\infty}^{\infty}d\omega K(\tau,\omega)A(\kv,\omega)
\end{align}
where $K = \frac{1}{\pi}e^{-\tau\omega}/(1+e^{-\beta\omega})$  and $G(\kv,\tau)$ is the single-particle Green's function. 

\begin{figure}[t]
    \centering
    \includegraphics[scale=0.4]{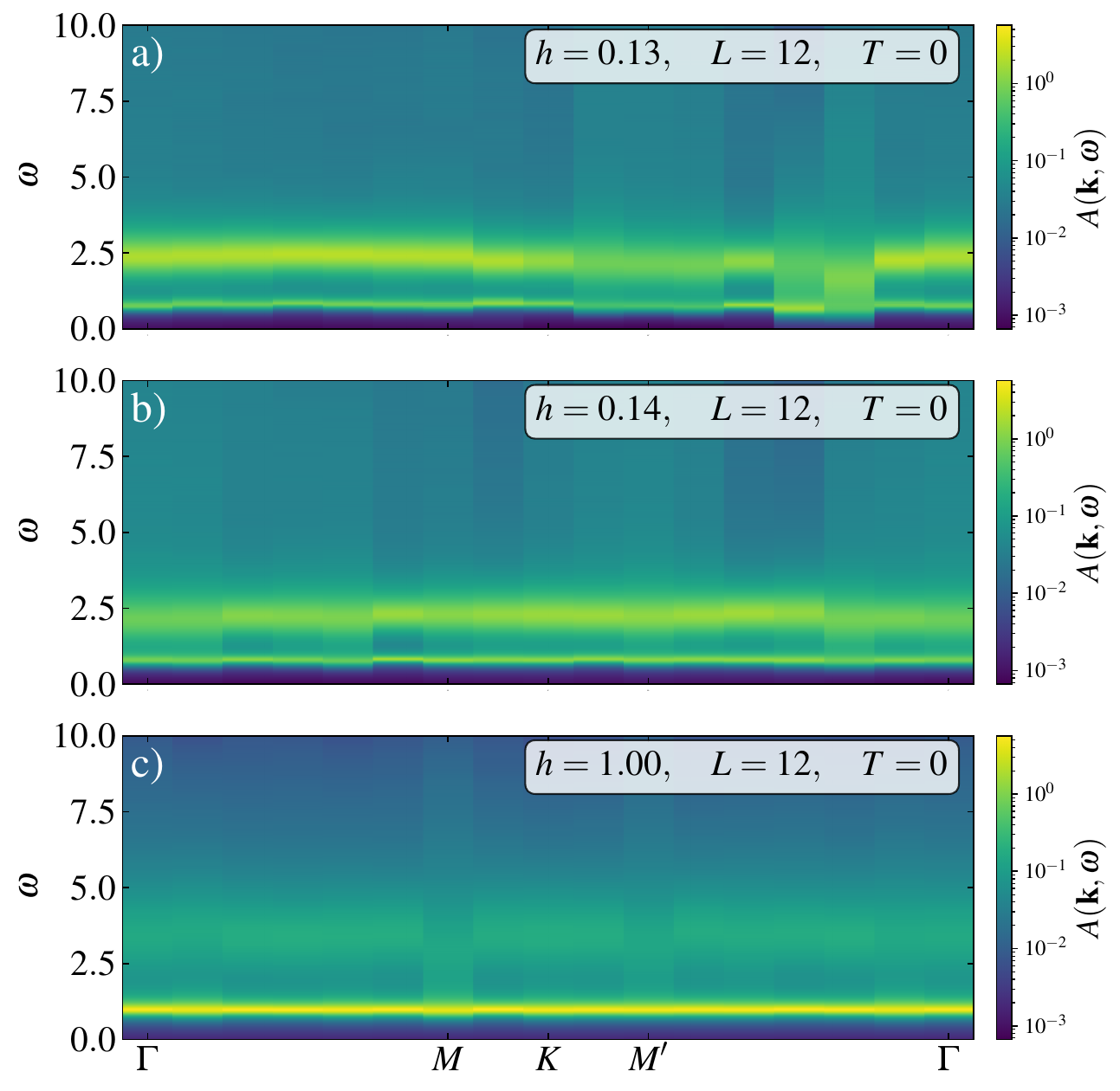}
    \caption{Single-particle spectral function $A(\kv,\omega)$ obtained from zero-temperature simulations of an $L=12$ triangular lattice at (a) $h=0.13$, (b) $h=0.14$, and (c) $h=1.00$. The spectra were obtained through the SAC method on the imaginary-time Green's function and are shown along the $\Gamma - M - K -M'$ path in the Brillouin zone. All panels share the same logarithmic color scale.}
    \label{fig:A_k_omega}
\end{figure}

We consider the single-particle spectral function, \(A(\mathbf{k},\omega)=-\mathrm{Im}\{G(\mathbf{k},\omega)\}\), across the two phases. Fig.~\ref{fig:A_k_omega} shows $A(\mathbf{k},\omega)$ for three representative values of the transverse field. Panels (a) and (b), corresponding to $h=0.13$ and $h=0.14$, respectively, lie within the vison crystal phase, whereas panel (c), at $h=1.00$, corresponds to the superfluid phase. A notable feature of these results is the absence of momentum dispersion, with spectral weight predominantly distributed along a flat band throughout the Brillouin zone. This behavior follows directly from the local $\mathbb{Z}_2$ gauge symmetry. As discussed in Sec.~\ref{subsec:symmetries}, the conservation of the $\hat{Q}_\iv$ operators enforces the spatial locality of the bare single-particle Green's function, \(G_{ij}(\tau)=\delta_{ij}G_{ii}(\tau)\). Furthermore, Elitzur's theorem states that a local symmetry cannot be spontaneously broken. Consequently, the spectral function satisfies $A(\mathbf{k},\omega)=A_{\mathrm{loc}}(\omega)$ for any value of $h$. Therefore, the absence of dispersion does not imply the absence of propagating fermionic excitations, but rather reflects the fact that the bare fermion operator carries a local $\mathbb{Z}_2$ gauge charge.

We also note that Fig.~\ref{fig:A_k_omega}  is consistent  with a gapped spectrum which  signals a gap  between the ground state  Gauss  sector  and  one  with a  single  flipped  gauge  charge.   This is consistent  with the notion  that the Gauss law is  dynamically  imposed in the   vison crystal  phase. Finally, it is worth mentioning that resolving the spectral functions becomes increasingly challenging at smaller transverse fields. Comparing Figs.~\ref{fig:A_k_omega}(a) and (b), the spectrum at $h=0.13$ exhibits stronger fluctuations than that at $h=0.14$. This highlights the numerical difficulties in the weak-field regime, where the gauge-field dynamics become progressively slower. Therefore, the residual momentum-dependent variations, particularly at $h=0.13$, should be interpreted with caution rather than as evidence of a genuine dispersion.

\begin{figure}[h!]
    \centering
    \includegraphics[scale=0.45]{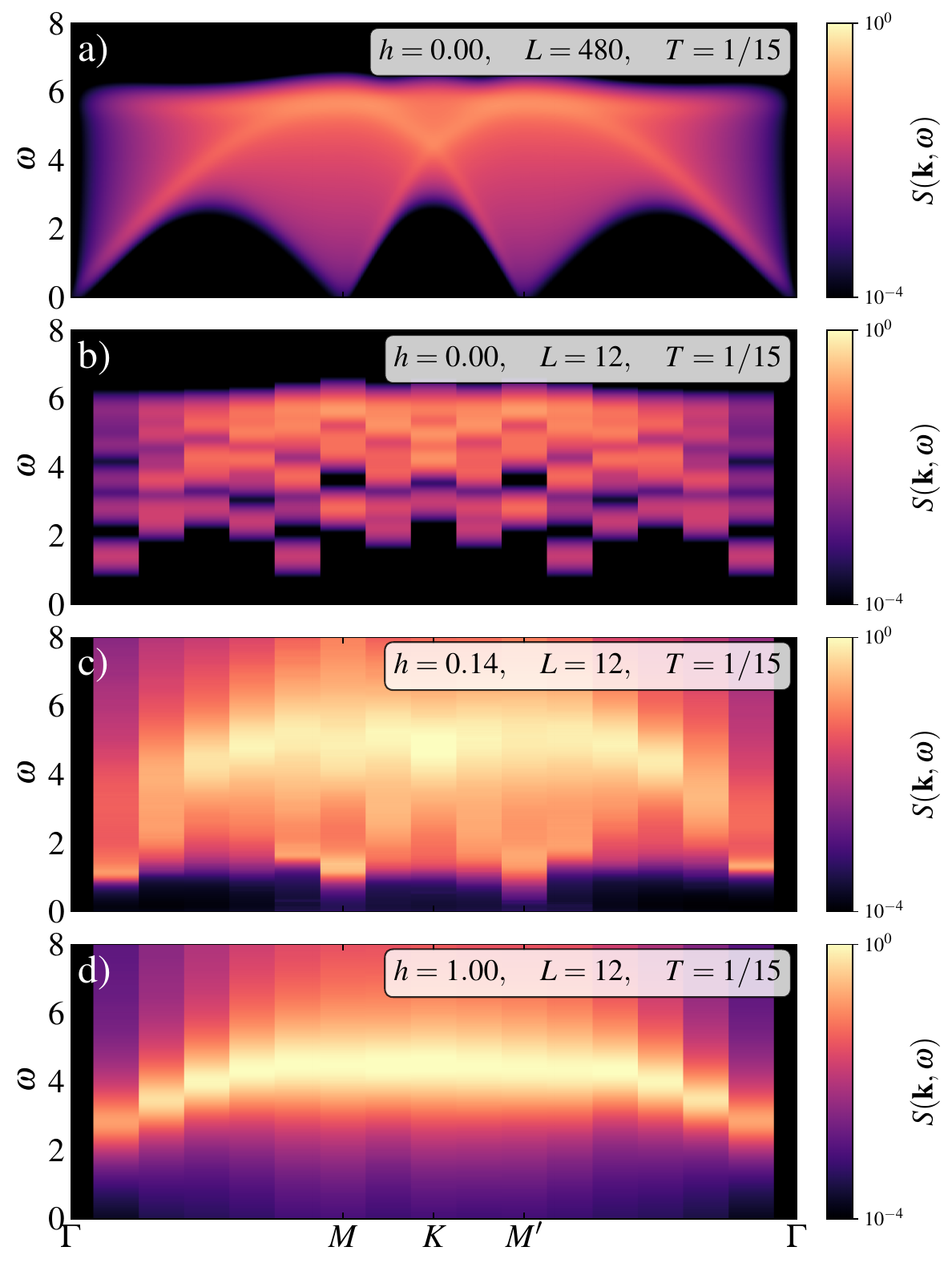}
    \caption{Dynamical spin susceptibility $S(\kv,\omega)$. (a), (b): fixed $\pi$-flux gauge at $h=0$ for $L=480$ and $L=12$ lattices, respectively. (c), (d): AF-QMC simulations of an $L=12$ triangular lattice at $\beta = 15$ for $h=0.14$ and $h=1.00$, respectively. The spectra were obtained through the SAC method on the imaginary-time spin correlation function and are shown along the $\Gamma - M - K -M'$ path in the Brillouin zone. All panels share the same logarithmic color scale.}
    \label{fig:S_k_omega}
\end{figure}

While the  $c$-fermions  remain local since they carry local $\mathbb{Z}_2$ gauge charge, the gauge-neutral  particle-hole spin excitations can propagate coherently.  In Fig.~\ref{fig:S_k_omega} we  compute the dynamical spin susceptibility at a temperature scale,  $T = 1/15$, that reflects ground state properties. In  Fig.~\ref{fig:S_k_omega}(a) we use a fixed gauge at $h=0$  and compute the aforementioned quantity in the extended zone scheme,  on an $L=480$ lattice.  This plot  serves as  a reference,  and  shows  gapless excitations connecting the two Dirac points, at $M$ and $M'$,  as  well as  within a single cone, at the $\Gamma$ point. The  envelope of the low-lying continuum shows a  linear dispersion.  Fig.~\ref{fig:S_k_omega}(b)  shows how finite  size  effects  manifest  themselves at $L=12$ still within the fixed gauge calculation in the  extended zone scheme. Although we observe a finite  size gap  at the $M$ and $M'$ points, we can still see that the spectral  function becomes soft at these points.  The  QMC simulation in the vison crystal phase at $h=0.14$, Fig.~\ref{fig:S_k_omega}(c), shows  clear  similarities with the  fixed  gauge, $h=0$ calculation  of Fig.~\ref{fig:S_k_omega}(b): in both cases we observe a softening of the spectral function at the $M$ and $M'$ points.  This provides numerical  support for the vison  crystal phase  described in Sec.~\ref{subsec:weak-coupling}.  In the $s$-wave SF  phase spin fluctuations are  gapped.  As apparent in Fig.~\ref{fig:S_k_omega}(d),  and in contrast to Fig.~\ref{fig:S_k_omega}(c), the  spectrum hardens.

\subsection{Finite temperature}
\label{subsec:FT}

\begin{figure}[h]
    \centering
    \includegraphics[scale=0.42]{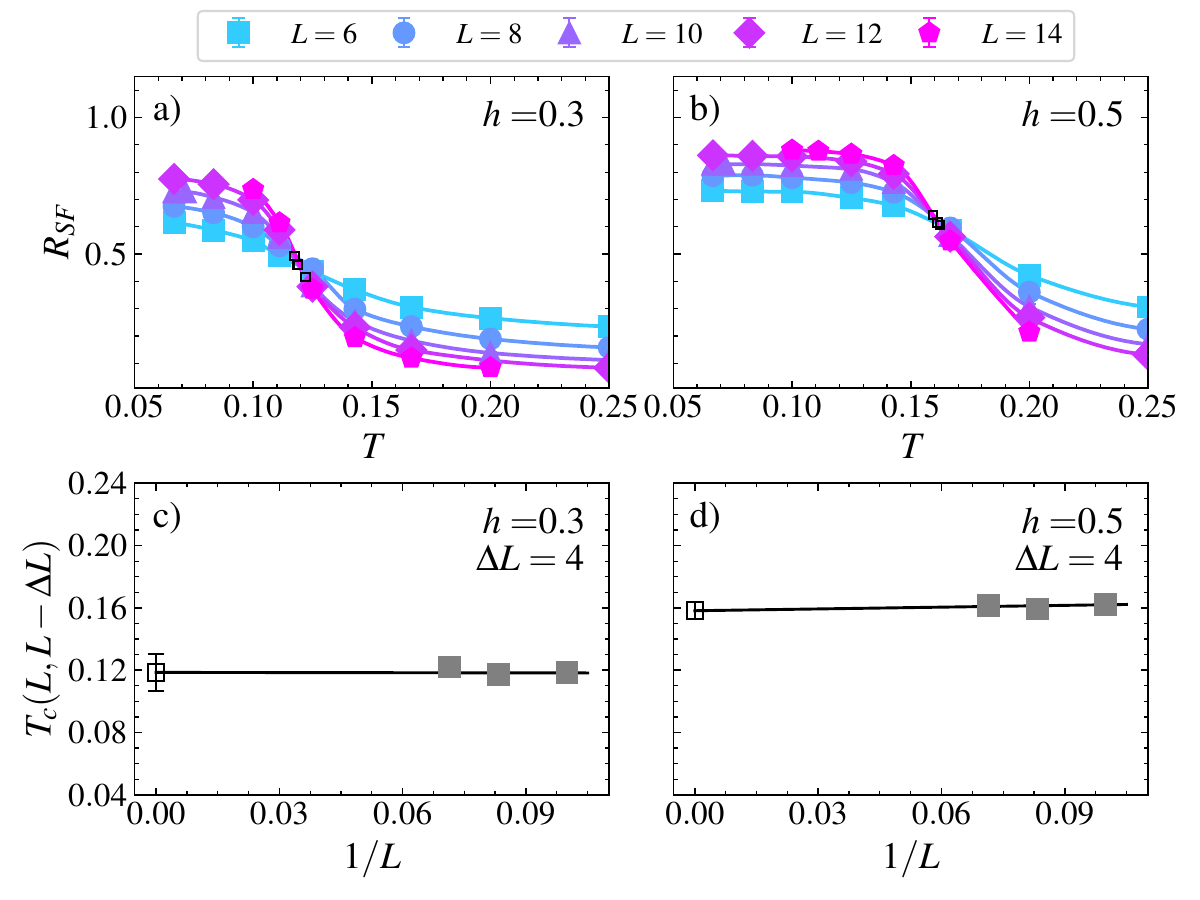}
    \caption{(a) Correlation ratio for the superfluid phase as a function of the temperature for $h=0.3$. Panel (b) is the same as panel (a) but for $h=0.5$. (c) Finite size scaling plot of the crossing points (filled squares) in panel (a) using sizes $L$ and $L-\Delta L$, with $\Delta L = 4$, leading to the extrapolated ($L\rightarrow \infty$) value $T_{KT} = 0.11(1)$ (empty square). (d) Same for the data in panel (b), leading to the extrapolated value $T_{KT} = 0.158(5)$.}
    \label{fig:R_sc_T}
\end{figure}

Having established the ground-state phase diagram, we next examine how stable the vison crystal and superfluid phases are against thermal fluctuations. As discussed before, these two phases break different symmetries: the former breaks the global $\mathbb{Z}_2$ symmetry, while the latter breaks the continuous U(1) symmetry. In two dimensions, this distinction leads to different thermal critical behaviors. For the vison crystal phase, the phase transition is expected to belong to the two-dimensional Ising universality class, while the superfluid phase is expected to melt through a Kosterlitz--Thouless (KT) transition. We investigate these transitions to determine the finite-temperature phase boundaries and construct the  $T \times h$ phase diagram shown in Fig.~\ref{fig:diagram}. We start with the superfluid thermal transition. To locate this transition, we perform a finite-size scaling analysis of the superfluid correlation ratio, which distinguishes the regime with robust superfluid correlations from the thermally disordered phase \cite{Fontenele2024}. Figure~\ref{fig:R_sc_T}(a)--(b) show typical plots for $R_{\mathrm{SF}}$ as a function of the temperature, for two representative values of the transverse field, $h=0.3$ and $h=0.5$, respectively. In both panels, $R_{\mathrm{SF}}$ is large at low temperatures and decreases upon heating, indicating the melting of superfluid correlations. The crossing points between curves for different lattice sizes provide finite-size estimates of the transition temperature ($T_{KT}$). We estimate the $T_{KT}$ in the thermodynamic limit by extrapolating these crossings using pairs of sizes $L$ and $L-\Delta L$, with $\Delta L=4$. As displayed in Fig.~\ref{fig:R_sc_T}(c)--(d), we obtain $T_{KT}=0.11(1)$ for $h=0.3$ and $T_{KT}=0.158(5)$ for $h=0.5$, respectively. Thus, the superfluid phase survives up to a finite temperature, and its transition temperature increases as $h$ is increased.

\begin{figure}[t]
    \centering
    \includegraphics[scale=0.45]{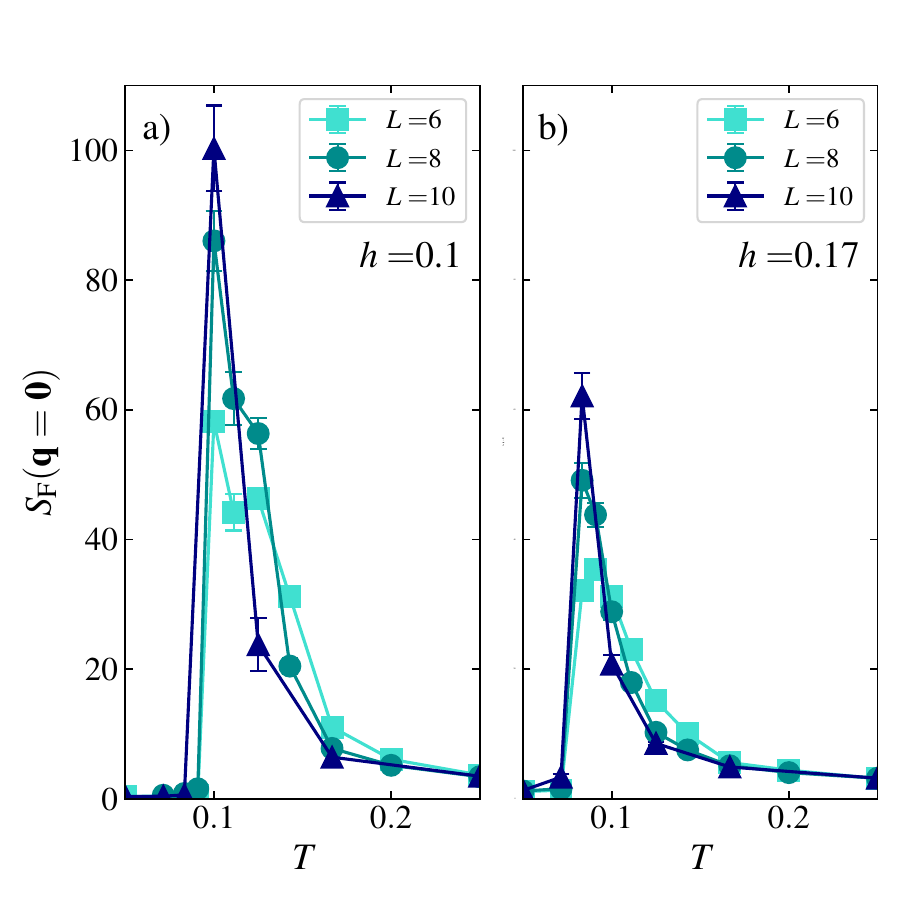}
    \caption{Flux structure factor at fixed (a) $h=0.1$ and (b) $h=0.17$ as a function of the temperature. The peak in the structure factors signals the critical temperature for the vison crystal phase at $T_c = 0.10(1)$ for the data in panel (a), and $T_c=0.08(1)$ for the data in panel (b). }
\label{fig:S_F_T}
\end{figure}

We further analyze the finite-temperature melting of the vison crystal phase. It is important to recall that, in contrast to the superfluid transition discussed above, the vison crystal phase is associated with the spontaneous breaking of the global $\mathbb{Z}_2$ symmetry introduced in Sec.~\ref{subsec:symmetries}. Since the vison crystal phase breaks the global $\mathbb{Z}_2$ symmetry, long-range order is allowed at finite temperature in two dimensions, and the continuous thermal transition is expected to belong to the two-dimensional Ising universality class \cite{Onsager1944, Pelissetto2002}. Although the thermal vison crystal transition belongs to a known universality class, accurately determining the critical points remains computationally demanding. This difficulty arises because Monte Carlo simulations near an Ising critical point are affected by the well-known problem of critical slowing down: the autocorrelation time increases significantly in the critical region, making it increasingly difficult to thermalize the simulations \cite{LandauBinder21}. For this reason, substantially longer simulations are required in the vicinity of the vison crystal phase transition. This computational limitation restricts our finite-temperature analysis to lattices up to $L=10$. Nevertheless, the finite-size dependence of the critical temperature ($T_c$) is weak. Therefore, these data are sufficient to provide reliable estimates of the critical temperature within our numerical accuracy.

\begin{figure}[t]
    \centering
    \includegraphics[scale=0.38]{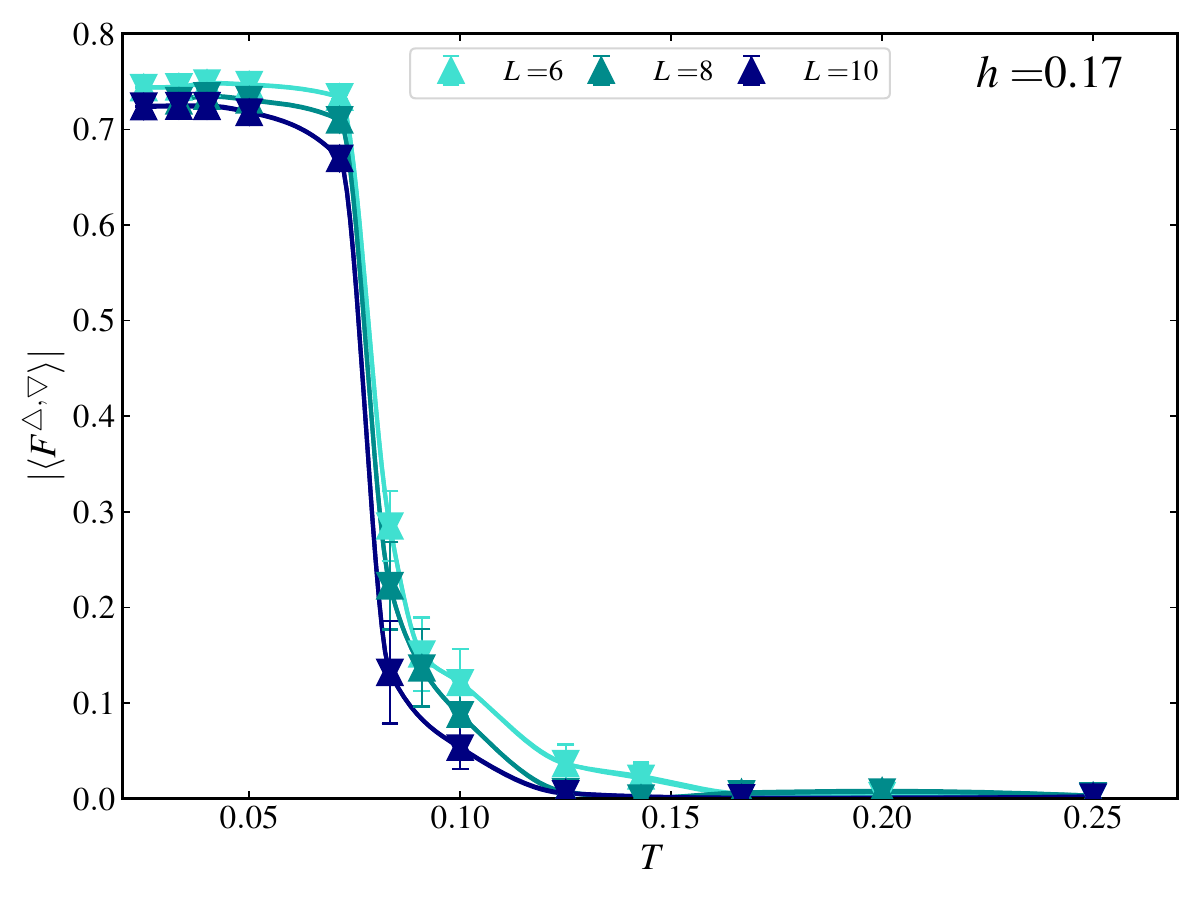}
    \caption{Absolute value of the average flux of the sublattices up (upward triangles) and down (downward triangles) as a function of the temperature for $h=0.17$. In contrast to the abrupt behavior observed in the ground state transition, the breaking of the global $\mathbb{Z}_2$ symmetry occurs continuously, as expected for a second-order phase transition. }
    \label{fig:flux_b}
\end{figure}

Flux fluctuations can be probed through the flux structure factor, with $S_{F}(\qv = 0)$ expected to develop a pronounced peak at the critical temperature. Figures~\ref{fig:S_F_T}(a)--(b) show the flux structure factor at $\qv=(0,0)$ as a function of temperature for two representative values of the transverse field, $h=0.1$ and $h=0.17$, respectively. For both values of $h$, $S_F(\mathbf q=0)$ develops a sharp peak at the critical temperature, signaling enhanced flux fluctuations associated with the restoration of the global Ising symmetry. From the position of these peaks, we estimate $T_c=0.10(1)$ for $h=0.1$ and $T_c=0.08(1)$ for $h=0.17$. This picture is also corroborated by the behavior of the absolute value of the average plaquette flux shown in Fig.~\ref{fig:flux_b}. In contrast to the abrupt jump observed at the ground-state transition, here  the absolute value of the average plaquette flux  decreases continuously as the temperature is increased and vanishes above the transition, consistent with a continuous Ising transition out of the vison crystal phase. At $h=0.17$,  consistency with the $T=0$  calculations  (see Fig.~\ref{fig:S_sc}) implies that there will be a temperature scale below which we observe a  transition out of the vison crystal phase to the  SF.  In Appendix~\ref{App:Clausius-Clapeyron} we provide an understanding of the slope of  the first-order transition between the vison crystal and superfluid phases, based on a  Clausius--Clapeyron relation.  The precise location of this phase boundary remains uncertain, and lies below temperatures accessible in our simulations.

\section{Conclusions}
\label{sec:conclusions}

In this work, we investigated an unconstrained $\mathbb{Z}_2$ gauge theory coupled to SU(2) fermions on the triangular lattice at half filling using sign-problem-free AF-QMC simulations. We showed that the frustrated non-bipartite geometry is responsible for profound changes in the phases and in the critical behavior of this model, as opposed to bipartite lattices \cite{Assaad2016}. In particular, the triangular geometry gives rise to a global $\mathbb{Z}_2$ symmetry that acts as an Ising symmetry on the plaquette fluxes. At small transverse field strength \(h\), it is spontaneously broken. We classify this phase as a vison crystal, i.e. a deconfined phase where the flux pattern breaks mirror symmetries,  and where gauge invariant quantities show Dirac-like signatures. At large $h$, quantum fluctuations restore the global $\mathbb{Z}_2$ symmetry, suppressing  flux order and stabilizing a superfluid phase with $s$-wave pairing. Since  $s$-wave pairs possess no gauge charge,  gauge  fluctuations decouple from the  fermionic matter in the SF phase. As such, in this phase,  the $\mathbb{Z}_2$  gauge theory can be  confined (SF) or deconfined (SF$^\star$).  In the large-$h$ limit, visons become  expensive and the gauge theory is confined. Since within the SF phase  we observe no  phase transitions in the flux structure factor, we  conclude  that  the gauge sector is  confined in the  SF  phase.
At zero temperature, these SF and vison crystal phases are separated by a direct first-order quantum phase transition at $h_c \simeq 0.15$. Across this  transition, the broken symmetry changes from the discrete global $\mathbb{Z}_2$ symmetry of the vison crystal phase to the continuous global U(1) symmetry of the superfluid and  at  which the gauge  field  undergoes a confinement transition.

We further extended our analysis to finite temperatures and determined the thermal phase boundaries of both phases. In the superfluid phase, the thermal transition is consistent with a Kosterlitz--Thouless transition, while in the vison crystal phase, we expect the transition to belong to the Ising universality class. For the SF phase, we estimated the transition temperatures through a thorough finite-size scaling analysis of the superfluid correlation ratio. For the vison crystal phase, we analyzed both the flux structure factors and the absolute value of the average plaquette flux to extract the critical temperatures. Combining these results, we obtained the $T\times h$ phase diagram, see Fig.~\ref{fig:diagram}. We also derived a Clausius--Clapeyron relation, see Appendix \ref{App:Clausius-Clapeyron}, for our model to justify the positive slope of the coexistence line in our phase diagram.


Connecting these results to parton constructions of spin Hamiltonians requires identifying the physical Gauss sector for the projection onto the spin Hilbert space.
Conservation of $\hat Q_\iv$ alone does not select this sector: the unconstrained model includes all configurations of the conserved charges. For the spin-$1/2$ fermionic parton interpretation considered here, the relevant sector is $Q_\iv=-1$, which reduces to the singly occupied spin subspace in the strong-field limit: $X_b=+1$. In contrast, our measurement at \(h = 4\) indicates the dynamical selection of the even $Q_\iv =+1$ sector. In the same strong-field limit, the even sector retains the empty and doubly occupied states, forming a charge pseudospin doublet, Eq.~\eqref{eq:ham_strong_h}. This connects the model to charge-sector parton constructions and the notion of a holon-doublon metal; in the strong-field regime studied here, these holon-doublon pseudospins develop superfluid order. The connection between charge and spin can be made explicit through a partial particle-hole transformation: $\hat c_{\iv,\uparrow} = \hat d_{\iv,\uparrow}$ and $\hat c_{\iv,\downarrow} = \hat d_{\iv,\downarrow}^{\dagger}$. It reverses the fermion parity and hence maps $Q_\iv \to - Q_\iv$. Thus, the superfluid maps to transverse ferromagnetic order. This transformation also reverses the hopping amplitude of one spin species, giving $t_\uparrow = -t_\downarrow$. On the triangular lattice, this relative sign cannot be removed by a sublattice transformation, so the mapping relates our results to a distinct model with spin-dependent hopping. 
Accessing the physical spin sector of the original Hamiltonian instead requires projection with $\hat P_{\mathrm{spin}}=\prod_\iv[(1 - \hat Q_\iv)/2]$. Implementing this projection in the present AF-QMC formulation will most likely result in a sign problem.

There are several possible extensions of this work. One possible route is to study a version of the \(\mathbb{Z}_2\) lattice gauge theory when the global \(\mathbb{Z}_2\) symmetry, \(\hat T\), is not present. For this one can introduce a term odd under $\hat T$, such as $- K \sum_\triangle \hat F_\triangle$, to the Hamiltonian which breaks this global symmetry while preserving the local conservation laws. Moreover, this term also supplies an independent magnetic energy scale and preserves the nonnegative weights of the original unconstrained simulation. This extension provides a way to test whether an independent flux energy scale stabilizes a more exotic deconfined phase without long range flux order and Dirac-like particle-hole excitations. Another possible extension is to exactly impose the constraint, restricting our simulations to one of the \(Q_\iv = \pm 1\) sectors. Restricting the sampling to the \(Q_\iv = -1\) sector would be particularly interesting due to the possible realisation of a \(\mathbb{Z}_2\) spin liquid \cite{shackleton_sign-problem-free_2025,yan_topological_2021}. Further directions include the generalization to SU(\(N_f\)) fermionic models, for which, at even \(N_f\), the AF-QMC formulation is free of the sign problem.

\section*{ACKNOWLEDGMENTS}
We thank Natanael C. Costa for discussions. We gratefully acknowledge the Gauss Centre for Supercomputing e.V. for funding this project by providing computing time on the GCS Supercomputer SUPERMUC-NG at Leibniz Supercomputing Centre (LRZ) (project number pn73xu) as
well as the scientific support and HPC resources pro-
vided by the Erlangen National High Performance Com-
puting Center (NHR@FAU) of the Friedrich-Alexander-
Universit\"at Erlangen-N\"urnberg (FAU) under the NHR
project b133ae. NHR funding is provided by federal
and Bavarian state authorities. NHR@FAU hardware
is partially funded by the German Research Foundation
(DFG) - 440719683). W.K. is grateful to the Brazilian Agencies Conselho Nacional de Desenvolvimento Cient\'\i fico e Tecnol\'ogico (CNPq) [Grants No.~201552/2025-6 and No.~140789/2025-1] and Coordena\c c\~ao de Aperfei\c coamento de Pessoal de Ensino Superior (CAPES). F.F.A  acknowledges financial support by the Deutsche Forschungsgemeinschaft (DFG, German Research Foundation) through the W\"urzburg-Dresden Cluster of Excellence {\it ctd.qmat} -- Complexity, Topology and Dynamics in Quantum Matter (EXC 2147, Project No.~390858490). 
J.C.I. acknowledges financial support by the Deutsche Forschungsgemeinschaft (DFG, German Research Foundation) through the grant 
AS 120/19-1, project number 530989922.

\appendix

\section{Clausius--Clapeyron relation}
\label{App:Clausius-Clapeyron}

In this Appendix, we derive the Clausius--Clapeyron relation used to determine the slope of the first-order transition line in the finite-temperature phase diagram. We consider a general free energy \(F(T,\lambda)\), where \(T\) is the temperature and \(\lambda\) is an external parameter that drives the transition. Its differential can be written as
\begin{equation}
    dF = -S\,dT + O\,d\lambda,
    \label{eq:appendix_free_energy_differential}
\end{equation}
where
\begin{equation}
    S = -\left(\frac{\partial F}{\partial T}\right)_\lambda, \qquad O = \left(\frac{\partial F}{\partial \lambda}\right)_T
    \label{eq:appendix_conjugate_variable}
\end{equation}
are the entropy and the thermodynamic variable conjugate to \(\lambda\), respectively.

Along a first-order coexistence line separating two phases, denoted by \(A\) and \(B\), the free energies are equal, \(F_A(T,\lambda)=F_B(T,\lambda)\). Therefore, infinitesimal displacements along the coexistence line must satisfy \(dF_A=dF_B\). Using Eq.~\eqref{eq:appendix_free_energy_differential}, we obtain
\begin{equation}
    -S_A\,dT + O_A\,d\lambda = -S_B\,dT + O_B\,d\lambda.
\end{equation}
Thus, we obtain the generalized Clausius--Clapeyron relation
\begin{equation}
    \frac{dT_c}{d\lambda} = \frac{\Delta O}{\Delta S}.
    \label{eq:appendix_general_CC}
\end{equation}

We now apply this result to our Hamiltonian, Eq.~\eqref{eq:hamiltonian}. The variable conjugate to $h$ is
\begin{equation}
    O_h = \left(\frac{\partial F}{\partial h}\right)_T = \left\langle \frac{\partial \hat{H}}{\partial h} \right\rangle,
    \label{eq:appendix_Oh_general}
\end{equation}
and hence,
\begin{equation}
    O_h = -\left\langle \sum_b \hat{X}_b \right\rangle.
    \label{eq:appendix_conjugate_h}
\end{equation}
Inserting Eq.~\eqref{eq:appendix_conjugate_h} into Eq.~\eqref{eq:appendix_general_CC}, we obtain
\begin{equation}
    \frac{dT_c}{dh} = -\frac{\Delta \left\langle \sum_b \hat{X}_b \right\rangle}{\Delta S}.
    \label{eq:appendix_CC_h_total}
\end{equation}
Equivalently, using intensive quantities,
\begin{equation}
    x = \frac{1}{N}\left\langle \sum_b \hat{X}_b \right\rangle, \qquad s = \frac{S}{N},
\end{equation}
where \(N\) is the number of lattice sites, Eq.~\eqref{eq:appendix_CC_h_total} becomes
\begin{equation}
    \frac{dT_c}{dh} = -\frac{\Delta x}{\Delta s}.
    \label{eq:appendix_CC_h_density}
\end{equation}

Now we can write 
\begin{equation}
    \Delta x = x_{\mathrm{SF}} - x_{\mathrm{VC}}, \qquad \Delta s = s_{\mathrm{SF}} - s_{\mathrm{VC}}.
\end{equation}
Since the expectation value of the transverse field is larger in the superfluid phase, \(x_{\mathrm{SF}}>x_{\mathrm{VC}}\), we have \(\Delta x>0\). On the other hand, the vison crystal phase is expected to have a larger entropy than the superfluid phase, \(s_{\mathrm{VC}}>s_{\mathrm{SF}}\), which implies \(\Delta s<0\). Therefore,
\begin{equation}
    \frac{dT_c}{dh} = -\frac{\Delta x}{\Delta s} > 0.
\end{equation}
Thus, the Clausius--Clapeyron relation implies that the first-order coexistence line has a positive slope in the $T\times h$ phase diagram.

\section{Auxiliary-field quantum Monte Carlo formulation}
\label{App:qmc_details}

In this Appendix, we outline the auxiliary-field quantum Monte Carlo formulation employed to simulate the dynamical $\mathbb Z_2$ gauge fields coupled to fermionic matter. We start by rewriting the Hamiltonian, Eq.~\eqref{eq:hamiltonian}, as
\begin{align}
    \hat{H} = \sum_b \hat{Z}_b \hat{K}_b - h\sum_b \hat{X}_b,
\end{align}
where $b=\langle i,j\rangle$ labels the bonds and
\begin{align}
    \hat{K}_b = \sum_{\sigma} \left( \hat{c}^{\dagger}_{i,\sigma}\hat{c}_{j,\sigma} + \mathrm{H.c.} \right)
\end{align}
is the fermion hopping operator. Working in the basis of the operator $\hat{Z}_b$, $\hat{Z}_b|s_b\rangle=s_b|s_b\rangle$ with $s_b=\pm1$, allows us to write the gauge fields as discrete Ising variables evolving along the imaginary-time direction. Using a Trotter decomposition with step size $\Delta\tau = \beta /L_{\tau}$, the imaginary-time axis is discretized and the quantum bond variables are represented by Ising configurations $s_b(\tau)$ defined on every bond and imaginary-time slice. The partition function then reads
\begin{align}
Z = \sum_{\{s_b(\tau)\}} C^{B L_\tau} & \exp \left[ J \sum_{\tau=1}^{L_\tau} \sum_b s_b(\tau+1)s_b(\tau) \right] \times \nonumber \\
& \times\mathrm{Tr}_{\mathcal{F}} \left[ \prod_{\tau=1}^{L_\tau} \prod_b e^{-\Delta\tau\,s_b(\tau)\hat{K}_b} \right],
\label{eq:qmc_partition}
\end{align}
where
\begin{align*}
    J = -\frac{1}{2}\ln\left[\tanh(\Delta\tau h)\right], \qquad
    C=e^{-J}\cosh{(\Delta\tau h)},
\end{align*}
$B$ is the number of bonds, $\mathrm{Tr}_{\mathcal{F}}$ means the trace over fermionic degrees of freedom and periodic boundary conditions in the imaginary-time axis, $s_b(L_\tau+1)=s_b(1)$, are assumed. Equation~\eqref{eq:qmc_partition} provides the form used in the Monte Carlo simulations. For a fixed space-time configuration $\{s_b(\tau)\}$, the fermionic part of the Hamiltonian is quadratic, so that the fermionic degrees of freedom can be exactly integrated out. Furthermore, since both spin flavors couple identically to the real-valued bond fields, their fermionic matrices are identical and real
\begin{align}
    M_\uparrow = M_\downarrow = M \in \mathbb{R}. \nonumber
\end{align} 
The resulting fermionic weight is therefore the square of a real determinant
\begin{align}
    W_F(\lbrace s_b \rbrace) = \text{det}[M_\uparrow]\text{det}[M_\downarrow] = (\text{det}[M])^2 \geq 0. 
\end{align}
Consequently, the model is free of the negative sign problem, even on non-bipartite lattices. It is worth emphasizing that, unlike auxiliary Hubbard–Stratonovich fields usually introduced to decouple quartic operators, the discrete fields $s_b(\tau)$ correspond directly to the physical $\mathbb{Z}_2$ degrees of freedom of the Hamiltonian.

For the ground state calculations with the projective AF-QMC, the thermal trace is replaced by the projection of a trial state $|\Psi_T\rangle$ with finite overlap with the ground state,
\begin{align}
\langle \hat{O}\rangle = \frac{\langle\Psi_T| e^{-\Theta\hat{H}/2} \hat{O} e^{-\Theta\hat{H}/2} |\Psi_T\rangle }{ \langle\Psi_T| e^{-\Theta\hat{H}} |\Psi_T\rangle },
\end{align}
where $\Theta$ controls the projection onto the ground state, which becomes exact in the limit $\Theta \rightarrow \infty$. Previous convergence tests have shown that choosing $\Theta = 2L$ yields systematic errors smaller than the statistical ones. Therefore, we adopt $\Theta = 2L$ throughout all projective AF-QMC calculations. For the  fermionic part of the trial wave function, $|\Psi_T^F\rangle$, we use the ground state of a noninteracting tight-binding Hamiltonian on the triangular lattice in the presence of a staggered $\pi$-flux background,
\begin{align}
    \hat{H}_{T} = \sum_{\langle i,j\rangle,\sigma} \eta_{ij} \left(\hat{c}_{i,\sigma}^{\dagger}\hat{c}_{j,\sigma} + \mathrm{H.c.} \right),
\end{align}
where $\eta_{ij}=\pm1$ is chosen such that the flux through neighboring plaquettes is always $\hat{F}_\rv = \pm 1$ in a staggered pattern. 

For the bosonic part of the trial wave function, $|\Psi_T^B\rangle$,  we set all the Ising spins in the x-direction: $\hat{X}_b|\Psi_T^B\rangle = |\Psi_T^B\rangle$  such that  in the computational basis where $\hat{Z}_b$ is diagonal all configurations on the  first and  last  time slices have the same weight.

\section{Low-energy expansion of the \(\pi\)-flux state}
\label{App:dirac_eq}

We consider a \(\pi\)-flux pattern on the triangular lattice in which all up triangles have a flux of \(\pi\) while all down triangles have zero flux. In this gauge configuration, the unit cell has two sites, belonging to two sublattices \(A\) and \(B\), and the lattice vectors are \(\mathbf{a}_1 = (2,0)\) and \(\mathbf{a}_2 = (1, \sqrt{3})/2\). The hopping Hamiltonian is then given by 
\begin{multline}
    \hat{H} = -t \sum_{i} \Big( \hat{c}_{\iv,A}^{\dagger} \hat{c}_{\iv,B}^{\phantom{\dagger}} - \hat{c}_{\iv,B}^{\dagger} \hat{c}_{\iv+\mathbf{a}_1,A}^{\phantom{\dagger}} + \hat{c}_{\iv,A}^{\dagger} \hat{c}_{\iv-\mathbf{a}_1+\mathbf{a}_2,B}^{\phantom{\dagger}} \\
    - \hat{c}_{\iv,A}^{\dagger} \hat{c}_{\iv+\mathbf{a}_2,A}^{\phantom{\dagger}} + \hat{c}_{\iv,B}^{\dagger} \hat{c}_{\iv+\mathbf{a}_2,A}^{\phantom{\dagger}} + 
    \hat{c}_{\iv,B}^{\dagger} \hat{c}_{\iv+\mathbf{a}_2,B}^{\phantom{\dagger}} + \text{H.c.} \Big).
\end{multline}
Doing a Fourier transform 
\begin{equation}
    \hat{c}_{\iv,\lambda}^{\phantom{\dagger}} = \frac{1}{\sqrt{N}} \sum_{\mathbf{k}} e^{i\mathbf{k}\cdot\iv} \hat{c}_{\mathbf{k},\lambda}^{\phantom{\dagger}},
\end{equation}
we arrive at 
\begin{equation}
    \hat{H} = \sum_{\mathbf{k}} 
    \begin{pmatrix} 
        \hat{c}_{\mathbf{k},A}^{\phantom{\dagger}} \\
        \hat{c}_{\mathbf{k},B}^{\phantom{\dagger}}
    \end{pmatrix}^\dagger
    \begin{pmatrix}
        \Delta_{\mathbf{k}} & S_{\mathbf{k}} \\
        S^*_{\mathbf{k}} & - \Delta_{\mathbf{k}}
    \end{pmatrix}
    \begin{pmatrix}
        \hat{c}_{\mathbf{k},A}^{\phantom{\dagger}} \\
        \hat{c}_{\mathbf{k},B}^{\phantom{\dagger}}
    \end{pmatrix}.
\end{equation}
Here \(\Delta_{\mathbf{k}} = 2t \cos(\mathbf{k}\cdot \mathbf{a}_2)\) and \(S_{\mathbf{k}} = -t \left( 1 + e^{-i \mathbf{k}\cdot(\mathbf{a}_2 - \mathbf{a}_1)} - e^{i \mathbf{k}\cdot\mathbf{a}_1} + e^{i\mathbf{k}\cdot \mathbf{a}_2} \right)\). The energy dispersion is given by \(\epsilon_{\mathbf{k},\pm} = \pm \sqrt{\Delta_{\mathbf{k}}^2 + |S_{\mathbf{k}}|^2}\). There are two gapless modes, with linear dispersion, at 
\begin{equation}
    \mathbf{K}_\pm = \left( 0, \pm \frac{\pi}{\sqrt{3}} \right).
\end{equation}

Expanding the Hamiltonian near the Dirac cones, 
\begin{equation}
    h_{\pm} = v_f \left( \gamma_\pm^x q_x + \gamma_\pm^y q_y \right),
\end{equation}
where \(v_f = \sqrt{6}t\) is the Fermi velocity and the Dirac matrices \(\gamma^\mu_\pm\) satisfy 
\begin{equation}
    (\gamma^\mu_\pm)^2 = 1, \quad \{\gamma^x_\pm, \gamma^y_\pm\} = 0,
\end{equation}
and are defined as 
\begin{equation}
    \gamma^x_\pm = \frac{1}{\sqrt{6}} (\mp \tau^x + 2 \tau^y \mp \tau^z ), \quad \gamma^y_\pm = \frac{\pm 1}{\sqrt{2}} (\tau^x - \tau^z).
\end{equation}

\section{Parton construction and mean-field theory}
\label{App:Mean-field}
In this appendix, we describe the orthogonal fermion mean-field theory for the Hamiltonian, 
\begin{equation}
    \hat{H} = \sum_{\ave{\iv,\jv}} \sum_{\s} \left( \hat{f}_{\iv,\s}^{\dagger}  
    \hat{\tau}_{\iv}^{z}\hat{Z}_{\ave{\iv,\jv}} \hat{\tau}_{\jv}^{z} 
    \hat{f}_{\jv,\s}^{\phantom{\dagger}}  + \text{H.c.} \right) -
    h \sum_{\ave{\iv,\jv}}\hat{X}_{\ave{\iv,\jv}}. 
\end{equation}
with the constraint, 
\begin{equation}
    \hat{g}_{\iv} = \hat{\tau}_{\iv}^{x} (-1)^{\hat{n}_{\iv}}  = 1.
\end{equation}

We  first show that  if  $\hat{Q}_{\iv} = \hat{\tau}_{\iv}^{x} \prod_{\delta} \hat{X}_{\ave{\iv,\iv+\delta}} =1$, then the parton Hamiltonian is equivalent to the original Hamiltonian.   Consider
\begin{equation} 
        \hat{G}_{\iv} = \hat{g}_{\iv} \hat{Q}_{\iv} = (-1)^{\hat{n}_{\iv}} \prod_{\delta} \hat{X}_{\ave{\iv,\iv+\delta}} 
\end{equation}    
and  the  canonical  transformations: 
\begin{equation}
        \hat{\tilde{Z}}_{\ave{\iv,\jv}} = \hat{\tau}_{\iv}^{z} \hat{Z}_{\ave{\iv,\jv}} \hat{\tau}_{\jv}^{z} 
\end{equation}
and 
\begin{equation}
        \hat{\tilde{X}}_{\ave{\iv,\jv}} = \hat{X}_{\ave{\iv,\jv}}.
\end{equation}
Then 
\begin{equation}
\hat{H}  = \sum_{\ave{\iv,\jv}} \sum_{\s} \left( \hat{f}_{\iv,\s}^{\dagger}  
\hat{\tilde{Z}}_{\ave{\iv,\jv}} 
\hat{f}_{\jv,\s}^{\phantom{\dagger}}  + \text{H.c.} \right) -
h \sum_{\ave{\iv,\jv}}\hat{\tilde{X}}_{\ave{\iv,\jv}}
\end{equation}
with constraint $\hat{G}_{\iv} = 1$.   This is nothing but the original model at $\hat{Q}_{\iv} = 1$. Essentially, due  to the double constraints  $\hat{Q}_{\iv} = 1$ and $\hat{g}_{\iv} = 1$, the  $\tau$ degrees of  freedom decouple from the Hamiltonian dynamics.

For the mean-field theory, we express the constraint as:
\begin{equation}
    \hat{\tau}_{\iv}^{x} =  \left(  1 - 2 \hat{n}_{\iv,\uparrow} \right) \left(  1 - 2 \hat{n}_{\iv,\downarrow} \right),
\end{equation}
and  impose  it  on average by including a Lagrange multiplier, $\lambda$,  in the mean-field Hamiltonian.  We furthermore  consider  the mean-field  parameters 
\begin{equation}
    e_{\ave{\iv,\jv}} = \langle \hat{\tau}_{\iv}^{z}\hat{Z}_{\ave{\iv,\jv}} \hat{\tau}_{\jv}^{z} \rangle, \quad 
    J_{\ave{\iv,\jv}} = \sum_{\s}\langle \hat{f}_{\iv,\s}^{\dagger} \hat{f}_{\jv,\s}^{\phantom{\dagger}}  + \text{H.c.} \rangle
\end{equation} 
and  $\langle \hat n_{\iv,\s} \rangle =  \frac{n}{2} $ with $n=1$ for the half-filled band. The resulting mean-field Hamiltonian reads:
\begin{eqnarray}
    & & \hat{H}_{\text{MF}} =   \sum_{\ave{\iv,\jv}} \sum_{\s} \left( \hat{f}_{\iv,\s}^{\dagger} e_{\ave{\iv,\jv}} \hat{f}_{\jv,\s}^{\phantom{\dagger}}  + \text{H.c.} \right) - h \sum_{\ave{\iv,\jv}}\hat{X}_{\ave{\iv,\jv}}  \nonumber \\ 
    & & + \lambda \sum_{\iv}   \hat{\tau}_{\iv}^{x}  + \sum_{\ave{\iv,\jv}} J_{\ave{\iv,\jv}} \hat{\tau}_{\iv}^{z}\hat{Z}_{\ave{\iv,\jv}} \hat{\tau}_{\jv}^{z}   - \sum_{\ave{\iv,\jv}} J_{\ave{\iv,\jv}} e_{\ave{\iv,\jv}}.
\end{eqnarray}
In this mean-field Hamiltonian, the bosonic and fermionic sectors decouple, and importantly, $\hat{Q}_{\iv} = \hat{\tau}_{\iv}^{x} \prod_{\delta} \hat{X}_{\ave{\iv,\iv+\delta}}$ remains locally conserved.   This implies that no assumption of the value of $\hat{Q}_{\iv}$ is required in the mean-field approximation.


\bibliography{ref,joao}

\end{document}